\documentclass[final,twocolumn,merge,sort&compress]{elsarticle}

\usepackage{color, subfigure, slashed, listings, fancyhdr, amsmath, amsthm, amssymb, bbm}
\usepackage{mathrsfs}
\usepackage{graphicx}
\usepackage{xcolor}

\usepackage[pdftex,
            plainpages=false,
            hypertexnames=true,
            pdfpagelabels=true,
            hyperindex=true,
            linktocpage,
            pagebackref=false,
            pdfa=true]{hyperref}

\usepackage{dcolumn}
\usepackage{xspace}
\usepackage{ulem}

\newcommand{\tr}{\ensuremath{\operatorname{tr}}}

\def\Eq#1{Eq.~(\ref{#1})}

\newcommand{\Tc}{\ensuremath{T_c}}

\definecolor{dblue}{rgb}{0,0,0.5}

\newcolumntype{C}[1]{>{\centering\hspace*{-3pt}}m{#1}}

\graphicspath{{../figures/}}

\begin{document}
\title{Center phase transition from matter propagators in (scalar) QCD}

\author[graz,BNL]{M. Mitter}

\author[graz]{M. Hopfer}

\author[giessen]{B.-J. Schaefer}

\author[graz]{R. Alkofer}

\address[graz]{Institut f\"ur Physik, NAWI Graz,
  Karl-Franzens-Universit\"at Graz, Austria}
\address[BNL]{Department of Physics, Brookhaven National Laboratory, Upton, NY 11973}
\address[giessen]{Institut f\"{u}r Theoretische Physik,
  Justus-Liebig-Universit\"{a}t Gie{\ss}en, Germany}

\date{\today}


\begin{abstract}
  Novel order parameters for the confinement-deconfinement phase
  transition of quenched QCD and fundamentally charged scalar QCD are
  presented. Similar to the well-known dual condensate, they are
  defined via generalized matter propagators with $U(1)$-valued
  boundary conditions. The order parameters are easily accessible with
  functional methods. Their validity and accessibility is explicitly
  demonstrated by numerical studies of the Dyson-Schwinger equations
  for the matter propagators. Even in the case of heavy scalar matter,
  where the propagator does not show a signature of the phase
  transition, a discontinuity due to the transition can be extracted
  in the order parameters, establishing also fundamentally charged
  scalar matter as a probe for color confinement.
\end{abstract}

\maketitle

\section{Introduction}
\label{sec:intro}

In the limit of infinitely heavy matter in the fundamental
representation of the gauge group, the deconfinement transition in
$SU(N)$ gauge theories can be characterized by the breaking and
restoration of center symmetry, cf.~e.g.~Ref.~\cite{PhysRevD.24.450}.
In the low temperature phase, purely gluonic systems are center
symmetric, and the phase transition to a deconfined state is related
to the spontaneous breaking of center symmetry. A corresponding order
parameter is given by the Polyakov loop
\begin{equation}
\label{eq:ploop}
L = \langle\text{\tr }\mathcal P \exp \int_0^\beta dx_0\,A_0(x)\rangle
\ , \quad \beta=1/T \ , 
\end{equation}
where the trace is taken over the gauge group algebra in the fundamental
representation, $\mathcal P$
denotes path ordering, and $A_0$ is the time-like component of the
gauge field degrees of freedom.  This quantity can be related to the
exponential of the negative free energy of a pair of free
fundamentally charged matter particles. As a consequence, a vanishing
Polyakov loop corresponds to an infinite energy cost for freeing the
matter particles, and is therefore interpreted as an indicator for a
confined matter phase.  This observation is not restricted to gauge
groups with a nontrivial center. Interestingly, for gauge groups with
a trivial center, such as the exceptional Lie group $G_2$, quenched
calculations reveal also a clear evidence for a first-order phase
transition \cite{Pepe:2006er, Greensite:2006sm, Cossu:2007dk,
  Danzer:2008bk, Maas:2012wr, Maas:2012ts}.  Although in this case the
Polyakov loop is not an order parameter in the strict sense, it is
still sensitive to the phase transition and behaves very similar to
its pendant in the $SU(3)$ case.

In the quenched limit of $SU(2)$, one finds a second-order phase
transition from a confined to a deconfined phase around a critical
temperature of $T_c\approx$ 230-240 MeV in continuum approaches like
the functional renormalization group (FRG) method or the variational
Hamiltonian approach \cite{Braun:2007bx, Fister:2013bh,
  Reinhardt:2016xci} whereas in numerical lattice simulations
$T_c \approx 295$ MeV \cite{Lucini:2012gg}. In contrast, for $SU(3)$, a
first order transition is found at $T_c\approx 275$ MeV in the FRG
\cite{Fister:2013bh}, at $T_c\approx 245$ MeV in the variational
Hamiltonian approach \cite{Reinhardt:2016xci}, and at $T_c\approx 270$
MeV on the lattice \cite{Lucini:2012gg}.

With functional methods like the Dyson-Schwinger equations (DSEs) and
the FRG, see e.g.\ \cite{Alkofer:2000wg, Roberts:2000aa,
  Fischer:2006ub, Aguilar:2008xm, Fischer:2008uz,
  Binosi:2009qm, Maas:2011se, Boucaud:2011ug, Eichmann:2016yit,
  Litim:1998nf, Berges:2000ew, Pawlowski:2005xe, Gies:2006wv,
  Schaefer:2006sr, Rosten:2010vm, Braun:2011pp, vonSmekal:2012vx} for
reviews and \cite{Braun:2007bx, Marhauser:2008fz, Fischer:2009wc,
  Braun:2009gm,
  Fischer:2009gk,Fischer:2010fx,Fischer:2011mz,Fischer:2012vc,
  Fister:2013bh, Haas:2013qwp, Fischer:2013eca, Fischer:2014ata,
  Quandt:2015aaa, Reinosa:2016iml, Huber:2016xbs, Cyrol:2017qkl} for
investigations of finite temperature Yang-Mills theory and the
center-symmetry transition, it is usually hard to access the Polyakov
loop directly, see also \cite{Herbst:2015ona}. Initiated by the
pioneering work on spectral sums in lattice gauge theory
\cite{Gattringer:2006ci, Bruckmann:2006kx, Bilgici:2008qy}, the
formulation of order parameters which are accessible to functional
methods, has been achieved in the last decade \cite{Synatschke:2007bz,
  Synatschke:2008yt, Fischer:2009wc}.  In particular, this concerns
the dual chiral condensate \cite{Fischer:2009wc}
\begin{eqnarray}
\label{eq:dualcond}
 \Sigma_1 & = &
 -\int\limits_{0}^{2\pi}\frac{d\varphi}{2\pi}e^{-i\varphi} \langle
 \bar \psi \psi \rangle_\varphi\ .
\end{eqnarray}
which has been successfully applied to the center phase transition in
quenched QCD \cite{Fischer:2009wc, Fischer:2009gk, Fischer:2010fx} as
well as in investigations of the QCD phase structure
\cite{Braun:2009gm, Fischer:2011mz, Fischer:2012vc,
  Fischer:2014ata}. It corresponds to a dressed Polyakov loop, i.e.\ a
sum over all loops that wind once around the torus
\cite{Bilgici:2008qy}. Therefore it transforms like the Polyakov loop
itself under center symmetry.

In this work, we introduce novel order parameters that allow to use
fundamentally charged matter as a probe for color confinement in the
quenched limit. In particular, we apply a new order parameter for QCD
that is based on a condensate, which remains finite even in the case
of non-vanishing current quark masses.  Furthermore, we investigate
scalar QCD, a QCD-like theory where the quarks are substituted by
fundamentally charged scalar fields, cf.\ \cite{Fister:2010ah,
  Fister:2010yw, Macher:2011ys, Huber:2011xc, Maas:2012tj,
  Mitter:2012tpo, Hopfer:2012qr, Mitter:2013me, Hopfer:2013via,
  Maas:2013aia, Maas:2014pba, Hopfer:2014szm,Maas:2016edk,
  Maas:2016ngo} and references therein.  It is expected that not only
quarks but any fundamentally charged matter is confined
\cite{Fister:2010yw}. To be more precise, the confined phase of scalar
QCD is actually continuously connected to a Higgs-like phase
\cite{Osterwalder:1977pc, Bertle:2003pj, Langfeld:2004vu}, however,
both ``phases'' can be distinguished by the infrared saturation of,
e.g., the gluon DSE \cite{Schaden:2013ffa}.  Nevertheless we will
treat (quenched) scalar QCD as if it would experience a confining
phase like (quenched) QCD and relate confinement to center symmetry.
An advantage of scalar QCD is a drastic simplification of the (finite
temperature) tensor structures in the higher $n$-point functions,
particularly in comparison to the quark-gluon vertex
\cite{Alkofer:2008tt, Williams:2014iea, Mitter:2014wpa,
  Williams:2015cvx, Binosi:2016wcx, Aguilar:2016lbe, Cyrol:2017ewj}.

This paper is organized as follows: We discuss the construction of
order parameters for center symmetry in
Sec.~\ref{sec:construction_op}.  The DSEs for (scalar) QCD and its
solution are discussed in Sec.~\ref{sec:DSE} and our numerical results
for the matter propagators and order parameters are presented in
Sec.~\ref{sec:centersym_nr}. We summarize our findings and conclude in
Sec.~\ref{sec:summary}.

\section{Center-symmetry order parameters}
\label{sec:construction_op}

In the following, we consider arbitrary matter fields $\Phi$ in the
fundamental representation of the gauge group. Applied to a gauge
theory at finite temperature with $\beta \equiv 1/T$, the center
symmetry transformations of the gauge fields $A_\mu = A_\mu^a T^a$ in
the adjoint representation of the gauge group $G$ and the fundamental
matter fields $\Phi$ read
\begin{eqnarray}
  \hspace{-0.6cm} \mathcal T_z A_\mu\left(\vec{x},x_4+\beta \right) \hspace{-0.2cm} & = & \hspace{-0.2cm} z A_\mu (\vec{x},x_4) z^\dag = A_\mu(\vec{x},x_4) \ , \label{eq:center_trafo_adj} \\
  \hspace{-0.6cm} \mathcal T_z \Phi\left(\vec{x},x_4+\beta\right) \hspace{-0.2cm} & = & \hspace{-0.2cm} z e^{i\varphi}\Phi(\vec{x},x_4) \ . \label{eq:center_trafo_fund} 
\end{eqnarray}
Here $z\in Z(G)$ labels the center elements of the gauge group $G$ and
$\mathcal T_z$ denotes a center-symmetry transformation. For the
$SU(3)$ gauge group, for example, the center elements are given by
$z \in \left\{0,e^{i2\pi/3},e^{i4\pi/3}\right\}$.  In
\Eq{eq:center_trafo_fund}, a general boundary condition
$e^{i\varphi}\in U(1)$ has been assumed for the matter field. Hence,
the gauge fields in the adjoint representation are invariant, while
the matter fields in the fundamental representation pick up a
non-trivial phase according to their boundary condition and are thus
not invariant under center transformations.

In analogy to Eq.~\eqref{eq:dualcond}, we define a general class of
order parameters~$\Sigma$ as the Fourier transform of a
condensate~$\Sigma_\varphi$, which depends on the generalized
$\varphi$-valued boundary conditions by
\begin{equation}
\label{eq:order_parameter_scalar}
 \Sigma = \int\limits_{0}^{2\pi}\frac{d\varphi}{2\pi}e^{-i\varphi} \Sigma_\varphi \ .
\end{equation}
The novel order parameters are sensitive to center-symmetry breaking,
transform as the conventional Polyakov loop under center
transformations, and can be easily extracted from the propagator DSEs.
To be an order parameter for the center phase transition, $\Sigma$
should transform under center transformation as
\begin{equation}
\label{eq:order_parameter_property}
\mathcal T_z \Sigma = z \Sigma\ .
\end{equation}
To achieve this general transformation property, it is sufficient that
the condensate~$\Sigma_\varphi$ fulfills
\begin{eqnarray}
  \Sigma_\varphi \hspace{-0.2cm} & = & \hspace{-0.2cm} \Sigma_{\varphi+2\pi} \ , \quad \varphi\in\left[0,2\pi\right[ \ , \label{eq:order_parameter_properties1} \\ 
  \mathcal T_z\Sigma_\varphi \hspace{-0.2cm} & = & \hspace{-0.2cm} \Sigma_{\varphi+\arg(z)} \ . \label{eq:order_parameter_properties2}
\end{eqnarray}
The first condition is the closure of the $U(1)$-valued boundary
condition and the second condition utilizes the unit modulus of the
center elements, which results in an additive shift of the boundary
angle by $\arg(z)$. It is obvious that
\Eq{eq:order_parameter_properties1} and
\Eq{eq:order_parameter_properties2} together with
\Eq{eq:order_parameter_scalar} implies
\Eq{eq:order_parameter_property}, i.e.\
\begin{equation}
\begin{split}\label{eq:order_parameter_proof}
  \mathcal T_z \Sigma & = \int\limits_{0}^{2\pi} \frac{d\varphi}{2\pi}
  e^{-i\varphi} \Sigma_{\varphi+\arg(z)} \\
  & = \int\limits_{\arg(z)}^{2\pi+\arg(z)} \frac{d\hat{\varphi}}{2\pi}
  e^{-i(\hat{\varphi}-\arg(z))} \Sigma_{\hat{\varphi}} = z \Sigma\ .
\end{split}
\end{equation}
Note that in the derivation of the order parameter no reference to the
ordinary Polyakov loop has been used.  Only the center-symmetry
transformation properties of the condensates, summarized in
Eqs.~\eqref{eq:order_parameter_properties1}
and~\eqref{eq:order_parameter_properties2}, together with the
integrability of the propagator dressing functions have been
exploited. The resulting order parameter transforms as the
conventional Polyakov loop, cf.~Ref.~\cite{Gattringer:2006ci}.  Thus,
the remaining task is the construction of condensates $\Sigma_\varphi$
that fulfill \Eq{eq:order_parameter_properties1} and
\Eq{eq:order_parameter_properties2}.

\subsection{Order parameter for scalar QCD}
\label{sec:application_sQCD}

The aim is now to find a condensate $\Sigma_\varphi$ that fulfills the
requirements of~\eqref{eq:order_parameter_properties1}
and~\eqref{eq:order_parameter_properties2} and can be expressed in
terms of the scalar propagator $D_{S,\varphi}$.  Applying
Eq.~\eqref{eq:center_trafo_fund}, we see that a center-symmetry
transformation changes only the boundary condition of the propagator
\begin{equation}
\label{eq:prop_centertrafo}
\mathcal T_z  D_{\varphi}(\vec{x},x_4) = D_{\varphi+\arg(z)}(\vec{x},x_4) \ .
\end{equation}
The sought for condensate $\Sigma_\varphi$ should therefore be related
to the propagator such that it depends only on the boundary condition
of the propagator.  Hence, we propose for real-valued propagators the
condensate
\begin{equation}
\label{eq:sigma_phi_1}
\Sigma_\varphi  \equiv \int\limits_{0}^\beta d\tau\, D^2_{\varphi} (\vec p=\vec 0, \tau) \ , 
\end{equation}
evaluated at vanishing three-momentum. This expression is obviously
periodic in $\varphi$ with periodicity $2\pi$, since $D_{\varphi}$ is
invariant under $\varphi \to \varphi+2\pi$.  The second requirement,
\eqref{eq:order_parameter_properties2}, is also trivially fulfilled
because of \eqref{eq:prop_centertrafo}.  Rather than evaluating the
propagator at some fixed $\tau$, or integrating over the propagator
itself, we choose to integrate over the square of the propagator in
\Eq{eq:sigma_phi_1}. This has the advantage that we can express
$\Sigma_\varphi$ as a sum over Matsubara modes of the momentum space
propagator
\begin{equation}
\label{eq:sigma_phi_2}
\Sigma_\varphi = T\sum\limits_{n} D^2_{\varphi} (\vec{p}=\vec 0, \omega_n(  \varphi )) \ ,
\end{equation}
with nice convergence properties.  From a mathematical point of view
this can be understood by analogy to the well-known $L^2$-norm.

\subsection{Order parameter for QCD}
\label{sec:application_QCD}

For the construction of a center-symmetry order parameter for QCD, we
start from the quark propagator
\begin{eqnarray}
\label{eq:quark_propagator}
  S(\vec{p},\omega_n) & = & \frac{-i\gamma^4 \omega_n\,C -i \vec{\gamma}\cdot\vec{p}\,A + B} {\omega_n^2 C^2 + \vec{p}^{\,2} A^2 + B^2}\ .
\end{eqnarray}
Here, the three dressing functions $A$, $B$ and $C$ depend on the
three-momentum $\vec p$ and the Matsubara mode $\omega_n$.  We propose
the generalized $\varphi$-dependent quark condensate
\begin{equation}
 \label{eq:order_parameter_quark}
 \Sigma_\varphi^{(q)} = T \sum\limits_n \left[\dfrac{1}{4}tr_D\;S(\vec{0},\omega_n(\varphi)) \right]^2,
\end{equation}
for the construction of a center-symmetry order parameter.  In
contrast to the conventional quark condensate
$\langle\bar\psi\psi\rangle_\varphi$ used in \Eq{eq:dualcond}, the
condensate $\Sigma_\varphi^{(q)}$ is finite even in the case of
non-vanishing bare quark masses. Furthermore, we expect a better
convergence compared to the dual scalar quark dressing, introduced in
Ref.~\cite{Fischer:2009gk}.

\section{Propagator Dyson-Schwinger equations}
\label{sec:DSE}

\subsection{Fundamentally charged scalar QCD}
\label{sec:scalarQCD}
\begin{figure*}[ht!]
  \centering 
  \subfigure{\includegraphics[width=0.8\linewidth]{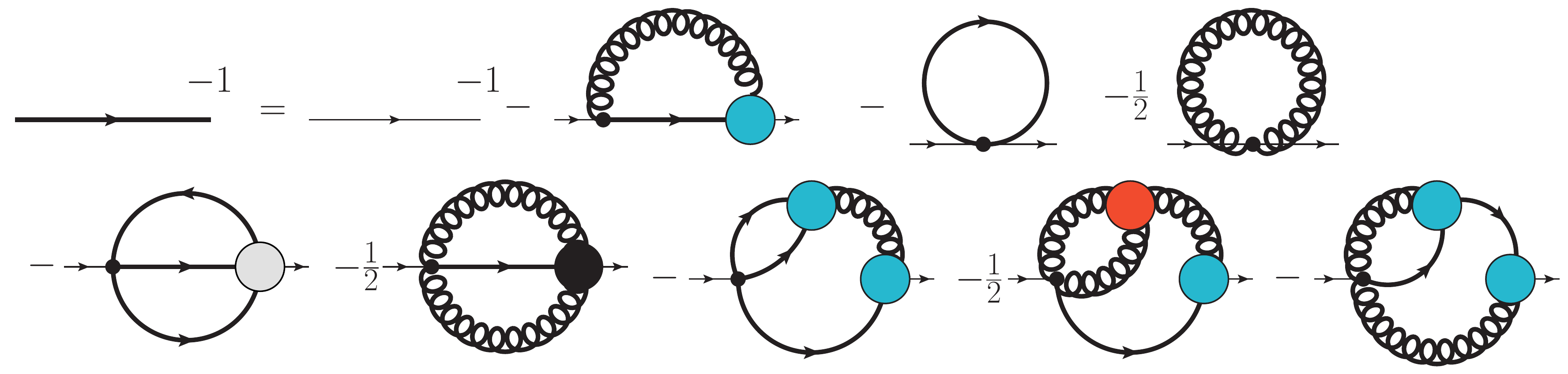}}
  \vspace*{-0.3cm}
  \caption{\label{fig:dse_scalprop} The DSE for the full scalar
    propagator includes full gluon propagators (curly lines), full
    (thick colored) and bare (small dots) vertices. See
    \cite{Fister:2010ah} for more details.  }
\end{figure*}

In QCD, the quark fields transform under the fundamental
representation of the $SU(3)$ gauge group. A
QCD-like theory that seems to be simpler at first glance can be
obtained by replacing the quark fields with complex scalar fields that
share the same transformation properties under the $SU(3)$ gauge
group. This leads to fundamentally charged scalar QCD whose
renormalized Euclidean Lagrangian in Landau
gauge~\cite{Fister:2010ah,Fister:2010yw} reads
\begin{equation}
\label{eq:action}
\begin{split}
& \mathcal L = Z_3A^a_\mu\left(\frac{1}{2}\left(-\partial^2 \delta_{\mu\nu}+\partial_\mu\partial_\nu\right)\right)A^a_\nu \\
 & + Z_1gf^{abc}\left(\partial_\mu A^a_\nu\right)A^{b}_\mu A^{c}_\nu 
   - \tilde{Z}_1g f^{abc}\bar{c}^a\partial_\mu(c^bA^c_\mu) \\
 & - \tilde{Z}_3\bar{c}^a\partial^2 c^a 
   + Z_4\frac{g^2}{4}f^{abc}f^{ade}A^b_\mu A^c_\nu A^{d}_\mu A^{e}_\nu \\
 & + ig\hat{Z}_{1F}T^a \phi^*\left(2A^a_\mu(\partial_\mu\phi)+(\partial_\mu A^a_\mu)\phi\right) \\
 & + \hat{Z}_4\frac{\lambda}{4}(\phi^*\phi)^2 + \phi^*\hat{Z}_2\left(-\partial^2+Z_m m^2\right)\phi \\
 & + \frac{\hat{Z}_{4,2}}{2}\{T^a,T^b\}g^2 \phi^* A^a_\mu A^b_\mu\phi \ .
\end{split}
\end{equation}
The gluon fields with the gauge coupling $g$ are denoted by $A^a_\mu$,
the (anti)ghosts fields by $(\bar{c}^a)$, $c^a$ and the fundamentally
charged scalar fields by $\phi$. Besides the coupling to the gauge
sector, renormalizability allows for a mass term $m$ and a quartic
coupling $\lambda$ in the scalar sector. Furthermore, the gauge
invariant coupling of the scalars to the glue sector implies the
presence of a two-scalar-two-gluon vertex. Compared to ordinary QCD,
scalar QCD has therefore two additional bare vertices, which results
in a richer diagrammatic structure of the corresponding DSEs. In particular,
this compensates to some extent the simplifications that are gained by replacing
Dirac spinors with scalar fields.
All primitively divergent vertices are endowed with renormalization
constants. Similarly to QCD, these are related by Slavnov-Taylor
identities like
$Z_3/Z_1 = \tilde{Z}_3/\tilde{Z}_1$~\cite{Marciano:1977su}, where
$\tilde{Z}_1=1$, and therefore $\hat{Z}_{1F} =\hat{Z}_2/\tilde{Z}_3$,
in Landau gauge \cite{Taylor:1971ff}.

\subsection{Scalar propagator Dyson-Schwinger equation\\}
\label{sec:DSE_scalprop}

The DSE for the inverse scalar propagator $D^{-1}_S(p)$ is depicted in
Fig.~\ref{fig:dse_scalprop}. Due to the presence of bare four-point
functions that involve the scalar field, its structure is considerably
more complicated than the corresponding DSE for the quark propagator.
Since our investigation focuses on the application of a new order
parameter for scalar QCD we use a truncated version of the propagator
equation in this work. The contribution from the additional two-loop
diagrams is suppressed in the coupling in the
ultraviolet. Furthermore, all but two of the two-loop diagrams are
additionally suppressed in the scalar mass, which is chosen to be of
order GeV. We found that too small masses
for the scalar field lead to instabilities in the numerical solution, most likely 
due to the vertex model or missing diagrams. For the purpose of this
work this large value for the mass is no restriction, since its main effect is the suppression
of matter backreactions, which is the definition of the quenched limit investigated
here.
Hence, major contributions from the two-loop diagrams,
which are neglected completely, are expected only in the mid-momentum
regime.  The main effect of the tadpole diagrams, on the other hand,
is a shift of the mass counterterm of the scalar propagator, which has
no significant impact on the results. Their non-perturbative
contribution is likely to be small compared to the error already
induced by the missing two-loop diagrams, see also
Ref.~\cite{Huber:2014tva} for an investigation of the tadpole diagram
in the gluon DSE. As a consequence, the tadpoles can safely be ignored
as well and the resulting DSE is of the same diagrammatic structure as
the one for the quark propagator.  The truncated scalar propagator DSE
is therefore finally given by
\begin{equation}
\label{eq:scalprop_dse}
\begin{split}
  & D_S^{-1} (p) = \hat{Z}_2\left(p^2+\hat{Z}_m m_0^2\right) - g^2C_F\hat{Z}_{1F} \\
  & \quad\times \int\frac{d^4q}{(2\pi)^4}\left(p+q\right)^\mu
  {D}^{\mu\nu}(k){D}_S(q) {\Gamma}^\nu(q,p) \ .
\end{split}
\end{equation}
with the Casimir invariant $C_F =(N_c^2-1)/(2N_c)$ and $k=p-q$.  The
dressed gluon (${D_{\mu\nu}}$) and scalar (${D}_S$) propagators are
parametrized by
\begin{equation}
\label{eq:prop_dressings}
 D_S(p) = \frac{Z_S(p^2)}{p^2} \ , \quad
 D^{\mu\nu}(p) =\frac{Z(p^2)}{p^2} P^{\mu\nu}(p) \ ,
\end{equation}
where the transverse projector
$P^{\mu\nu}(p) = \delta^{\mu\nu}-{p^\mu p^\nu}/{p^2}$ has been
introduced.  The reduced scalar-gluon vertex ${\Gamma}^\mu(p,q)$ is
defined by
\begin{equation}
\label{eq:scalar_gluon_vertex}
 {\Gamma}_a^\mu(p, q, k) =
 gT_a\left(2\pi\right)^4\delta^{(4)}\left(p+k-q\right){\Gamma}^\mu(p,q) \ .
\end{equation}
The general ansatz for the reduced vertex function contains two
dressing functions $A'$ and $B'$ \cite{Ball:1980ay, Ball:1980ax} 
\begin{equation}
 \label{eq:scalgluevertex_dressing}
\begin{split}
 {\Gamma}^\mu &(p,q) = A'(p^2,q^2;\xi) (p+q)^\mu \\
& +  B'(p^2,q^2;\xi)\left( p^\mu \left[q^2-p\cdot q\right]+q^\mu \left[p^2-p\cdot q\right]\right)
\end{split}
\end{equation}
with the abbreviation $\xi = p\cdot q/\sqrt{p^2 q^2}$.
With these definitions the DSE for the scalar propagator,
\Eq{eq:scalprop_dse}, can be expressed in terms of the dressing
functions
\begin{equation}
 \label{eq:scaldress_dse}
\begin{split}
  Z^{-1}_S &(x) = \hat{Z}_2\left(1+\hat{Z}_m \frac{m_0^2}{x}\right) \\
 & - \hat{Z}_{1F}C_F\alpha(\mu)\frac{2}{\pi^2}\int_0^\infty
 \!\!\!\!dy\, y\;Z_S(y)\;\mathcal{K}(x,y) \ ,
\end{split} 
\end{equation}
where
\begin{equation}
 \label{eq:scaldress_dse_2}
\begin{split}
  \mathcal{K}(x,y) = \int_{-1}^{1}\!\!\!
  &d\xi\left(1-\xi^2\right)^{3/2}\frac{Z(z)}{z} \\ 
 & \times\left\{\frac{A'(y,x;\xi)}{z}+\frac{B'(y,x;\xi)}{2}\right\} \ .
\end{split} 
\end{equation}
Here, $x=p^2$, $y=q^2$ and $z=x+y-2\sqrt{xy}\,\xi \equiv k^2$ denote the
squared external and internal scalar and gluon momenta, respectively.
The renormalized gauge coupling $\alpha(\mu)=g^2(\mu)/(4\pi)$
is evaluated at the renormalization scale $\mu$.

\begin{table*}[t]
\centering
\begin{tabular}{C{0.7cm}||C{0.4cm}|C{0.5cm}|C{0.4cm}|C{0.5cm}|C{0.55cm}|C{0.55cm}|C{0.55cm}|C{0.55cm}|C{0.55cm}|C{0.55cm}|C{0.35cm}|C{0.35cm}|C{0.35cm}|C{0.35cm}|C{0.35cm}|>{\hspace*{-3pt}}c}
\hline\hline
$T/T_c$      &0	  & 0.361& 0.44 & 0.451 & 0.549 & 0.603 &0.733  &0.903	&0.968 	&0.986 	&1 	&1.02	& 1.04 	&1.1 	&1.81 	&2.2\\\hline\hline
$a_{L}(T)$   &0.60& 0.42&0.23	& 0.33 & 0.19	&0.17	&0.11	&0.098	&0.082	&0.079	&0.16	&0.27	&0.32	&0.50	&2.71	&4.72\\
$b_{L}(T)$   &1.36& 1.23&1.14	& 1.20 & 1.13	&1.08	&1.10	&1.13	&1.14	&1.14	&1.05	&1.05	&1.03	&1.07	&1.14	&1.47\\\hline
$a_{T}(T)$   &0.60& 0.71&0.78	& 0.83 & 0.86	&1.04	&1.05	&1.67	&1.57	&1.06	&0.54	&0.55	&0.57	&0.63	&1.47	&1.42\\
$b_{T}(T)$   &1.36& 1.37&1.46	& 1.47 & 1.52	&1.60	&1.60	&1.91	&1.81	&1.45	&1.13	&1.14	&1.17	&1.19	&1.49	&1.30\\\hline\hline
\end{tabular}
\caption{The fit parameters for the temperature dependent SU(3) gluon
  propagator, adopted from \cite{Fischer:2010fx}.}
\label{tab:finiteT_gluon_fit}
\end{table*}

The generalization of the scalar propagator DSE to finite temperatures
is done within the Matsubara formalism with generalized $U(1)$-valued
boundary conditions, where the fourth momentum component is replaced
by discrete Matsubara frequencies
\begin{equation}
p_4\rightarrow \omega_n (\varphi)= (2\pi n +\varphi )T \text { with }
n \in \mathbb{Z}\ .
\end{equation}
Consequently, the momentum integral in the $p_4$ direction is
replaced by a sum over the discrete Matsubara modes. For periodic
boundary conditions, i.e.\ $\varphi = 0$, bosonic Matsubara
frequencies are used. Since the heat bath explicitly breaks Lorentz
symmetry, the scalar propagator dressing is now a function of the
spatial momentum $\vec{p}$ as well as the Matsubara frequency
$\omega_n$,
\begin{eqnarray}
\label{eq:prop_dressings_T}
  D_S(\vec{p}, \omega_n) & = & \frac{Z_S(\vec{p}^{\,2},\omega_n)}
  {\omega_n^2 + \vec{p}^{\,2}} \ .
\end{eqnarray}
The gluon propagator dressing splits into an electric ($L$) and a
magnetic ($T$) component, which are longitudinal and transverse to the
heat bath,
\begin{equation}
\label{eq:gluonprop_dressings_T}
D^{\mu\nu}(p) = P^{\mu\nu}_L(p)\frac{Z_L(p)}{p^2} + P^{\mu\nu}_T (p)\frac{Z_T(p)}{p^2} \ ,
\end{equation}
with the projection operators $P^{\mu\nu}_L = P^{\mu\nu}-P^{\mu\nu}_T$
and
$P^{\mu\nu}_T(p) = \delta^{i\mu}\delta^{i\nu} \left(\delta^{ij}-{p^i
    p^j}/{\vec{p}^{\,2}}\right)$.  Although the vertex would admit a
similar splitting, we will ignore this possibility for simplicity and
employ a degenerate vertex dressing for the electric and magnetic
components of the scalar-gluon vertex. Since we are using a model
for the vertex anyway, we will use the model parameter to fix its overall
effect to the physically required result. The consequences of choosing a vertex
model are dicussed further in the results section. Thus, the finite temperature
DSE can immediately be inferred from~\Eq{eq:scalprop_dse}, leading to
\begin{equation}
\label{eq:scaldress_dse_T}
\begin{split}
  \hat{Z}_2^{-1} &Z_S^{-1}(x,\omega_m) =
  1+\hat{Z}_m\frac{m^2}{\omega_m^2+x}
  -\frac{C_F \alpha(\mu)}{2\pi} \\
  & \times T\sum\limits_{n\in\mathbb{Z}}\int_{0}^\infty \!\! dy\sqrt{y}\frac{Z_S(y,\omega_n)}{\omega_n^2+y} \mathcal{A}(x,\omega_m,y,\omega_n)  \ , \\
\end{split}
\end{equation}
with
\begin{equation}
\label{eq:scaldress_dse_T_2}
\begin{split}
  & \mathcal{A}(x,\omega_m,y,\omega_n) 
   =
    \frac{4}{\omega_m^2+x}\int_{-1}^1 d\xi \Bigg\{\frac{Z_L\left(k^2\right)}{k^2} \\
  & \times \frac{\left(\omega_m^2+x\right) \left(\omega_n^2+y\right) -
      \left(\omega_m\omega_n+\sqrt{xy}\,\xi\right)^2}{k^2} \\
  & +
    \frac{Z_T\left(k^2\right)-Z_L\left(k^2\right)}{k^2}
    \frac{xy\left(1-\xi^2\right)}{\vec{k}^2}\Bigg\} \\
  & \hspace{2cm} \times A(x,\omega_m^2,y,\omega_n^2;\xi) \ ,
\end{split}
\end{equation}
where $\vec{k}^2 = x+y-2\sqrt{xy}\,\xi$ are now the corresponding
spatial momenta and $k^2 = (\omega_m-\omega_n)^2 + \vec{k}^2$.

\subsection{Gluon propagator}
\label{sec:gluon_input}

One important ingredient for the numerical treatment of the matter
propagator DSEs is the temperature-dependent gluon propagator. A
self-consistent incorporation of the corresponding Yang-Mills DSEs is
beyond the scope of this work. Therefore, we proceed analogously to
previous works, see e.g.~Refs.~\cite{Fischer:2009wc, Fischer:2009gk,
  Fischer:2010fx}, and use lattice results to fit the gluon
propagator.  For the two dressings in \Eq{eq:gluonprop_dressings_T},
the following fit function is employed~\cite{Fischer:2010fx}
\begin{equation}
\begin{split}
\label{eq:finiteT_gluon_fit}
Z_{T,L}(q,T) & = \frac{q^2 \Lambda^2}{(q^2 + \Lambda^2)^2} \Bigg\{\left(\frac{c}{q^2 + a_{T,L}\Lambda^2} \right)^{b_{T,L}} \\
& \hspace{-0.5cm} +
\frac{q^2}{\Lambda^2}\left(\frac{\beta_0\alpha(\mu)\ln\left[q^2/\Lambda^2
      + 1\right]}{4\pi} \right)^\gamma\Bigg\} \ .
\end{split}
\end{equation}
Here $c=11.5$ GeV$^2$ and the anomalous dimension is $\gamma=-13/22$.
The renormalization scale is fixed to $\alpha(\mu) = 0.3$ and
$\beta_0 = 11N_c/3$ is the universal $\beta$-function coefficient of
the quenched theory. The temperature-independent scale parameter
$\Lambda=1.4$ GeV reflects the gapping scale of the gauge sector of
the theory. The temperature-dependent parameters $a_{T,L}$ and
$b_{T,L}$ for the magnetic and electric dressing functions are used to
fit the different temperature dependence in the magnetic and electric
gluon propagators and listed in Tab.~\ref{tab:finiteT_gluon_fit}.

\begin{figure*}
  \centering
  \subfigure{\includegraphics[width=0.48\linewidth]{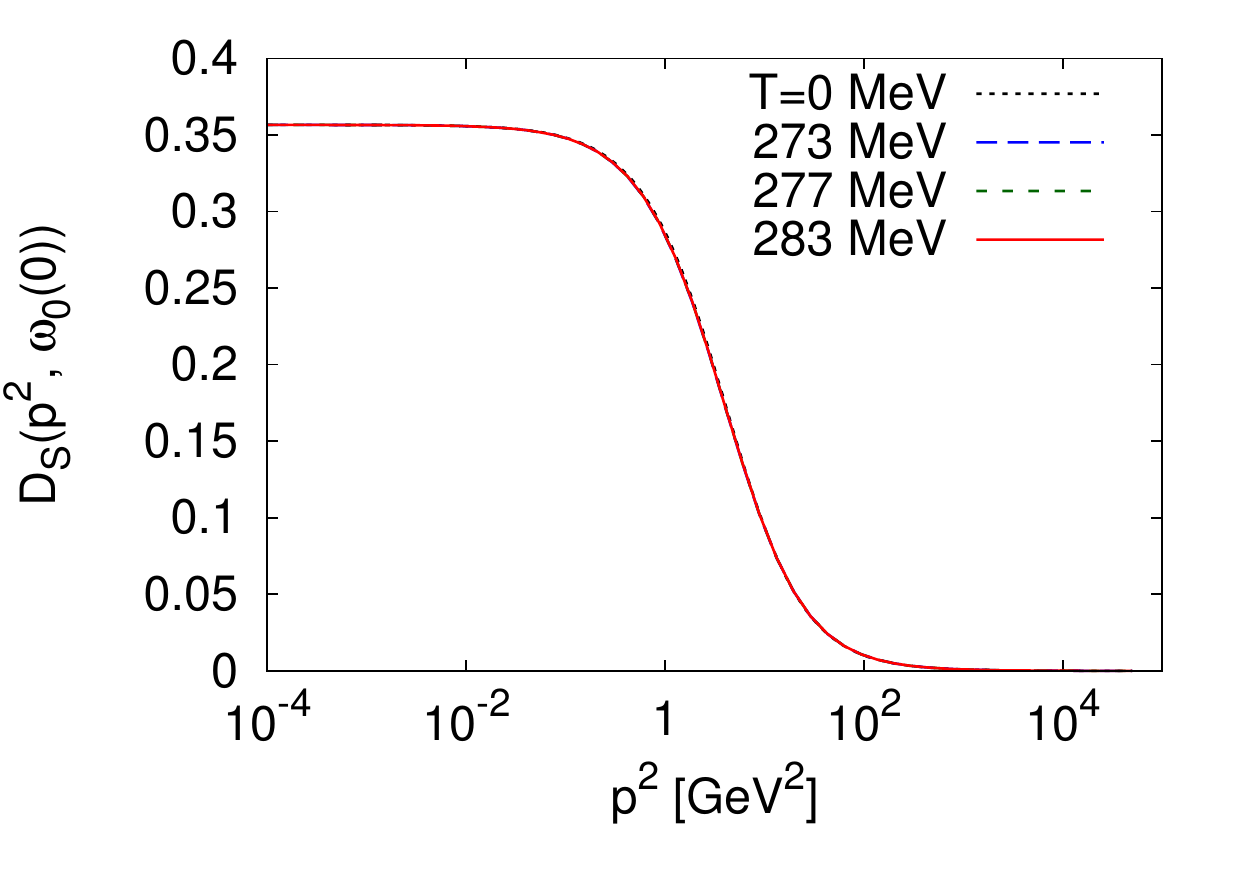} \label{fig:dse_scalprop_1_T_periodic}}
  \hfill
  \subfigure{\includegraphics[width=0.48\linewidth]{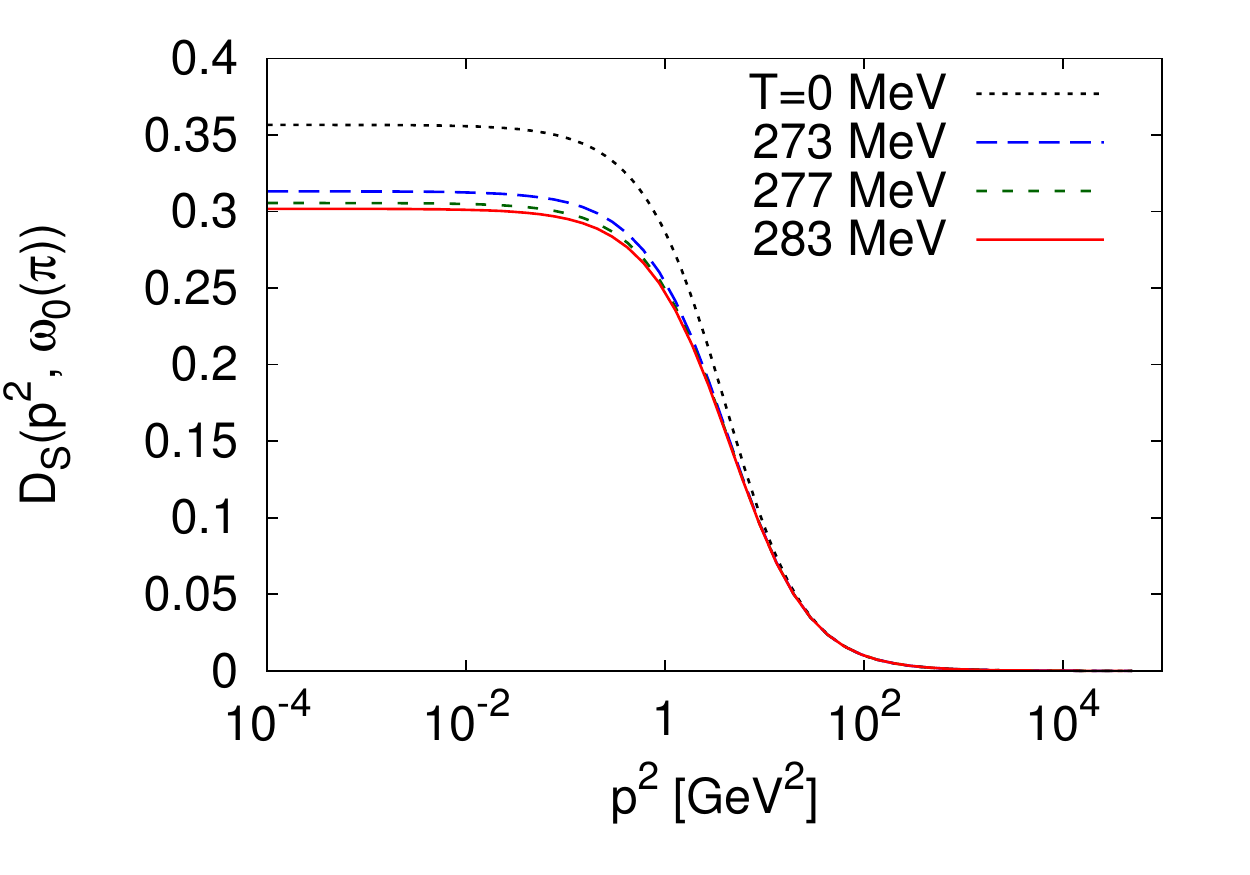} \label{fig:dse_scalprop_1_T_antiperiodic}}
  \caption{\label{fig:scalarprop} Scalar propagator,
    Eq.~\eqref{eq:scaldress_dse_T}, for periodic $\omega_0 (0)= 0$
    (left) and for antiperiodic $\omega_0(\pi)=\pi T$ (right) boundary
    conditions as a function of momenta for different temperatures
    ($\mu = 4$ GeV and $m = 1.5$ GeV).}
\end{figure*}

\subsection{Scalar-gluon vertex}
\label{sec:vertex}

For the complete numerical solution of the Dyson-Schwinger equation
for the scalar propagator the scalar-gluon vertex is also needed.
Since the DSE for the scalar propagator, \Eq{eq:scaldress_dse},
depends only on the sum of the two dressing functions that define the
scalar-gluon vertex, \Eq{eq:scalgluevertex_dressing}, it is sufficient
to provide only one model dressing function. The construction of the
model dressing functions $A$ is done analogous to the very successful
model dressing functions used for the quark-gluon vertex in
investigations of the QCD phase structure~\cite{Fischer:2010fx} and
reads explicitly
\begin{eqnarray}
\label{eq:scalgluevert_model}
 && \hspace{-0.75cm} A(x,y;\xi) = \tilde{Z}_3\, \frac{D_S^{-1}(x)-D_S^{-1}(y)}{x-y}\, d_1\times 
  \\ && \hspace{-0.75cm} \Bigg\{\!\!\left(\frac{\Lambda^2}{\Lambda^2 + z}\right) 
   + \frac{z\, \beta_0}{\Lambda^2 + z} \left(\frac{\alpha(\mu)
      \ln\left[\frac{z}{\Lambda^2} + 1\right]}{4\pi} \right)^{2\delta}\Bigg\} \ . \nonumber
\end{eqnarray}
Again, $x=p^2$, $y=q^2$ and $z=x+y-2\sqrt{xy}\,\xi \equiv k^2$ denote
the squared external and internal scalar and gluon momenta,
respectively.  The multiplication with the propagator-dependent
function is derived from Ward identities of scalar electrodynamics and
represents a generalization of the Ball-Chiu
vertex~\cite{Ball:1980ay, Ball:1980ax}.  Together with the anomalous dimension
$2\delta=-18/44$ and the values of $\beta_0$ and $\alpha(\mu)$ defined
in the previous section, it guarantees the one-loop consistent running
of the model dressing function in the perturbative ultraviolet regime
\cite{Fischer:2003rp}. The remaining part of the dressing function is
purely phenomenological. In particular, the parameter $d_1=0.53$
is introduced to model vertex effects that keep the 
order parameter as close to zero as possible, as required in the confining phase
below the transition temperature. 
The scale $\Lambda$ models the transition to the
nonperturbative regime of the theory, characterised by the gapping
scale of the glue sector. Finally, the vertex is multiplied with
$\tilde{Z}_3$ in order to replace $\hat{Z}_{1F}$ with $\hat{Z}_2$ in
\Eq{eq:scaldress_dse}.

At finite temperature we use the same model for the vertex with
\begin{eqnarray}
\label{eq:scalgluevert_model_finitet}
 A(x,\omega_m^2,y,\omega_n^2;\xi) = A(x+\omega_m^2,y+\omega_n^2;\xi)\ . \nonumber
\end{eqnarray}
In order to avoid singularities that could arise due to numerical
inaccuracies in the case
\begin{equation}
 \begin{split}
 \label{eq:singular}
  & \omega_m^2 + \vec{p}^{\,2} = \omega_n^2 + \vec{q}^{\,2}\ ,\quad \vec{q}\neq \vec{p}\ ,\\
  & D_S(\vec{p}^{\,2},\omega_m)\neq D_S(\vec{q}^{\,2},\omega_n)\ ,
 \end{split}
\end{equation}
the vacuum propagator in \Eq{eq:scalgluevert_model} is used also at
finite temperature.

\subsection{Renormalization}
\label{sec:renormalisation}

\begin{figure*}
  \centering
  \subfigure{\includegraphics[width=0.48\linewidth]{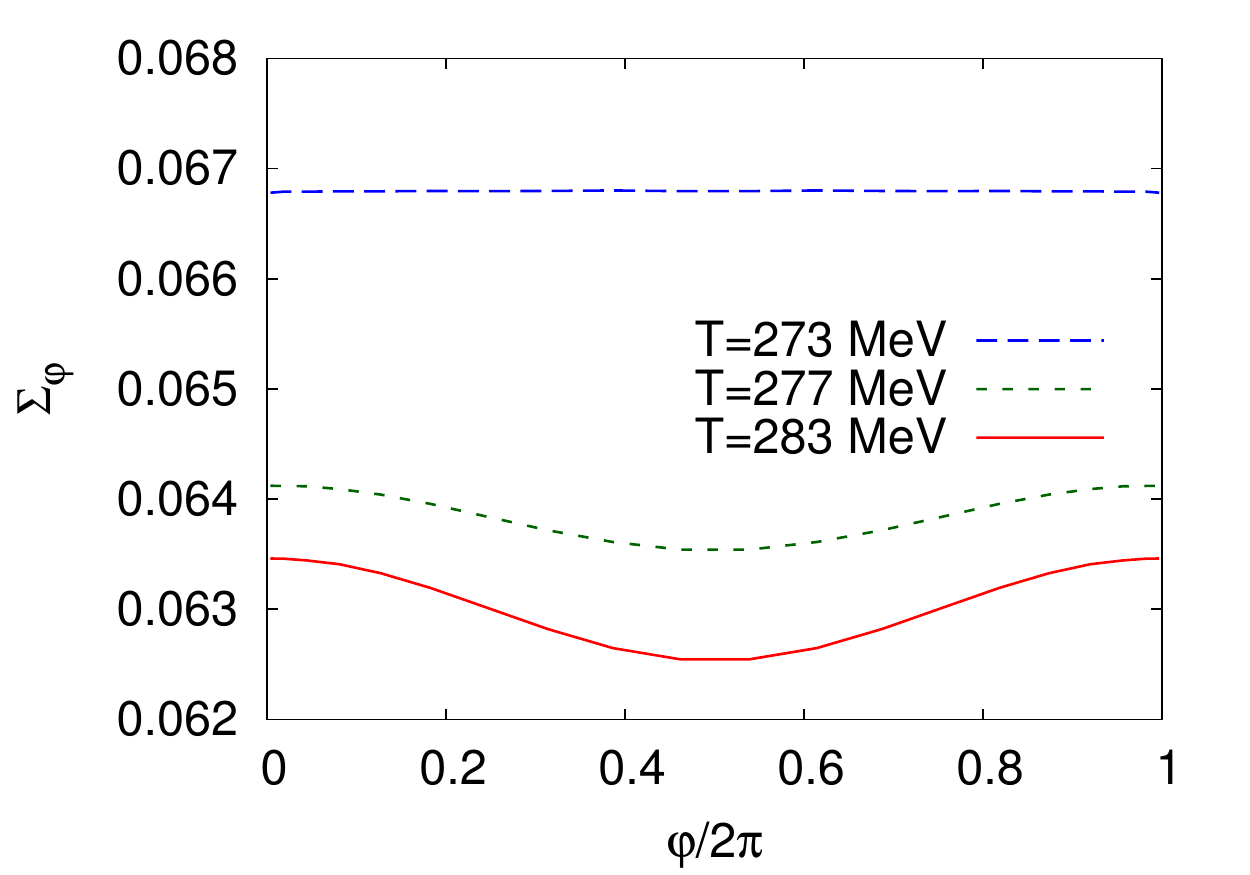} \label{fig:dse_dual_phi_1}}
  \hfill
  \subfigure{\includegraphics[width=0.48\linewidth]{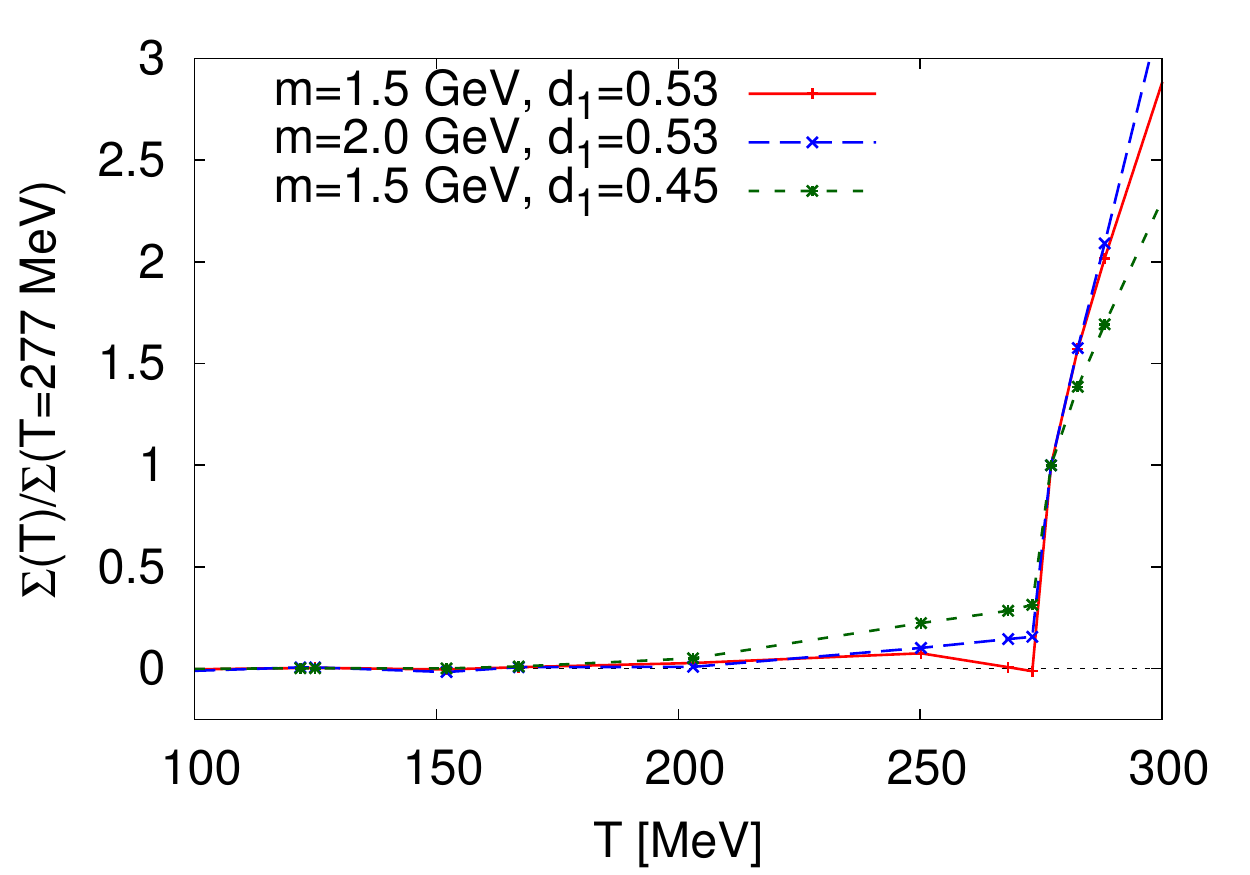} \label{fig:dse_dual_phi_2}}
  \caption{\label{fig:scalar_condensate} Left panel: Scalar function
    $\Sigma_\varphi$, Eq.~\eqref{eq:sigma_phi_2}, as a function of the
    boundary conditions for different temperatures ($\mu = 4$ GeV and
    $m = 1.5$ GeV). Right panel: Normalized scalar condensate as a
    function of the temperature for different scalar masses $m$ and
    fit parameter $d_1$ for the scalar-gluon vertex
    \Eq{eq:scalgluevert_model}.}
\end{figure*}

At finite temperature, as well as in the vacuum, the divergent
self-energy term is regularized with a sharp cutoff, where
a four-dimensional cutoff is used, i.e.
$x \leq \Lambda_c^2$ in the vacuum and $\omega_n^2 + x \leq \Lambda_c^2$
at finite temperatures with
$\Lambda_c^2 = 5\times 10^4$ GeV$^2$. We choose the renormalization
scale and conditions such that the counterterms remain unaffected by
the introduction of temperature. The self-energy term does not
contribute to $\hat{Z}_m m_0^2$ at vanishing momentum and no quadratic
divergencies are present in our truncation. Hence, for the
determination of the renormalization constant $\hat{Z}_m$ we fix the
value of the propagator at vanishing momentum via the
(cutoff-dependent) condition
$m^2 \equiv [ \hat{Z}_{1F} D_S(0) ]^{-1} = \hat{Z}_m m_0^2$.  In order
to fix $\hat{Z}_2$, we demand for $\mu^2>\Lambda_c^2$ that the
propagator is proportional to the free massless one,
$D_S(\mu^2) = 1/\mu^2$, which in turn yields
$\hat{Z}_2 = \mu^2 \left(\mu^2 + \mu^2\Pi(\mu^2) + m^2\right)^{-1}$.
This prescription together with the vertex in
(\ref{eq:scalgluevert_model}) ensures multiplicative renormalizability
in the sense that
$\hat{Z}_2(\mu^2,\Lambda_c^2)Z_S\left(x,\mu^2\right) =
\hat{Z}_2(\tilde\mu^2,\Lambda_c^2)Z_S\left(x,\tilde\mu^2\right)$ for
another scale $\tilde{\mu}$.

\subsection{Quark propagator Dyson-Schwinger equation}
\label{sec:DSE_quarkprop}

In addition to the center-symmetry order parameters calculated with
the scalar propagator in scalar QCD, we also present results for the
corresponding order parameters in QCD with quarks. The DSE for the
quark propagator and its solution have been discussed in great detail
in many previous works, see in particular \cite{Fischer:2010fx}. Since
we evaluate the order parameters on top of the results for the quark
propagator already presented therein, we refrain from repeating the
corresponding details.

\section{Numerical results}
\label{sec:centersym_nr}

We begin our investigation with the numerical solution of the scalar
propagator DSE~\eqref{eq:scaldress_dse_T}, using a
temperature-dependent gluon propagator from lattice simulations as
input, cf.\ Sec.~\ref{sec:gluon_input}.

In Fig.~\ref{fig:scalarprop} the scalar dressing function $D_S$,
Eq.~\eqref{eq:prop_dressings_T}, is shown as a function of the
momentum for different temperatures. The left panel shows the
propagator at the lowest periodic Matsubara frequency
$\omega_{n=0} (0)= 0$ and the right panel for the lowest antiperiodic
Matsubara frequency $\omega_{n=0}(\pi) = \pi T$.  In the case of
periodic boundary conditions, the scalar propagator is virtually
temperature independent, even above the phase transition temperature
$\Tc=277$ MeV.  For antiperiodic boundary conditions, on the other
hand, a trivial effect is induced by the temperature-dependent lowest
Matsubara frequency $\omega_{n=0} (\pi)= \pi T$.  Independent of the
boundary conditions, it is therefore hard to extract any useful
information about the transition temperature from the propagator
directly. This temperature insensitivity can be explained by the
relatively large scalar mass of $m \sim \mathcal{O}(1)$ GeV in
comparison to the transition temperature $T \sim \mathcal{O}(0.1)$
GeV.  Consequently, the scalar field is too heavy to be excited by the
fluctuations near the phase transition.

We show the condensate $\Sigma_\varphi$, defined in
Eq.~\eqref{eq:sigma_phi_2}, in Fig.~\ref{fig:scalar_condensate}, left
panel, and find immediately that it is indeed periodic in the
$U(1)$-valued boundary conditions.  Below the transition temperature,
$\Sigma_{\varphi}$ is $\varphi$-independent which in turn results in a
vanishing of the order parameter $\Sigma$. Above the transition
temperature, on the other hand, $\Sigma_\varphi$ develops a minimum
around $\varphi=\pi$.  This leads to a lowered negative contribution
in $\Sigma$, yielding ultimately an overall non-vanishing positive
value for the order parameter. For periodic boundary conditions, we
also find that $\Sigma_{\varphi=0}$ rises after a small dip above
$T_c$ rather quickly with the temperature. On the other hand, the
large-temperature growth of $\Sigma_{\varphi=\pi}$ is more moderate,
leading to an overall increase of the order parameter. In general,
this behaviour is similar to the one found in the literature, e.g.,
with spectral sums of the Dirac-Wilson operator
\cite{Synatschke:2007bz}, the dual quark condensate evaluated on the
lattice \cite{Bilgici:2008qy} or with functional methods
\cite{Fischer:2009wc,Fischer:2009gk,Braun:2009gm,Fischer:2010fx}.  In
the right panel of Fig.~\ref{fig:scalar_condensate}, the scalar order
parameter is displayed as a function of temperature for two different
scalar masses $m$ and different model parameters $d_1$ for the
scalar-gluon vertex. We observe small fluctuations of
the order parameter about the required zero value below the transition temperature.
These fluctuations are most likely caused by statistical 
errors in the lattice data. Additionally, we see that, depending on the 
parameter $d_1$ in the vertex model, a small but visible unphysical growth or decline
in the order parameter can be induced below the critical temperature, where
the discontinuity of the phase transition occurs. Nevertheless,
we observe a distinct phase transition, which
could not be seen in the propagator itself.  We conclude that
\Eq{eq:order_parameter_scalar} with (\ref{eq:sigma_phi_2}) is a valid
order parameter construction. In contrast, we could not find a linear
propagator formulation which emphasizes the importance of the
$L^2$-convergence property once more.

\begin{figure*}
  \centering
  \subfigure{\label{fig:dse_dual_quark_1}\includegraphics[width=0.43\linewidth]{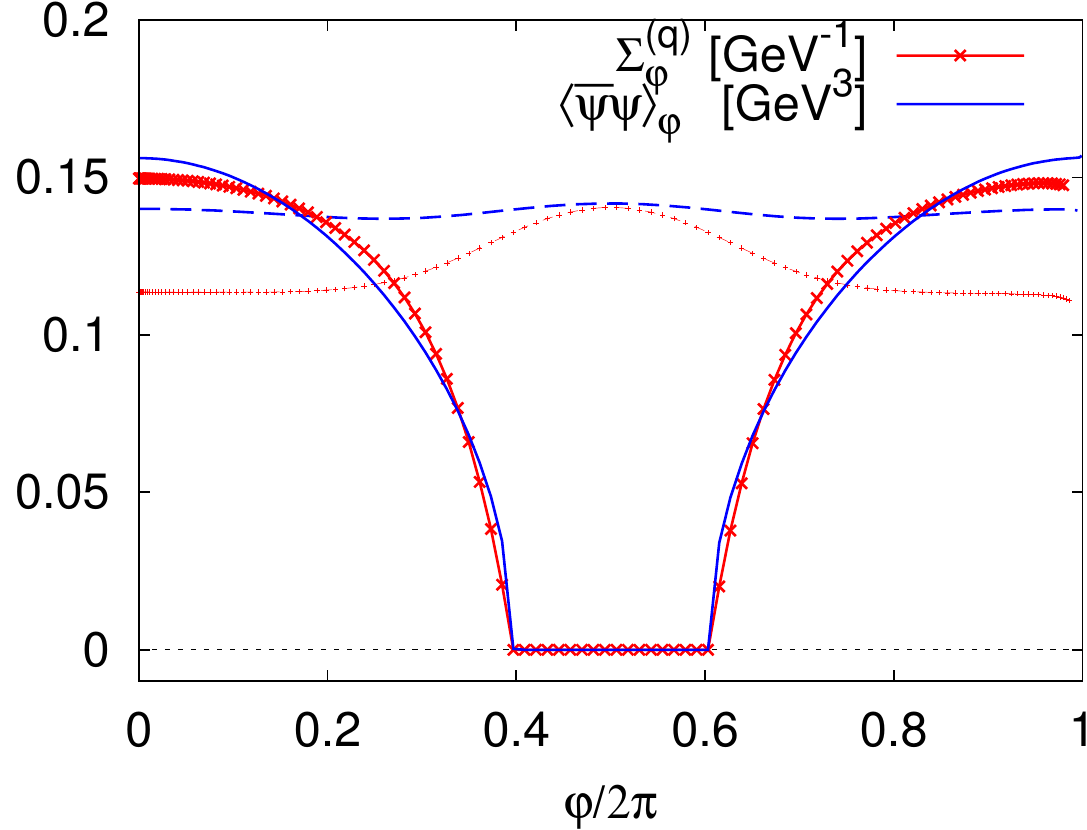}}
  \hfill
  \subfigure{\label{fig:dse_dual_quark_2}\includegraphics[width=0.43\linewidth]{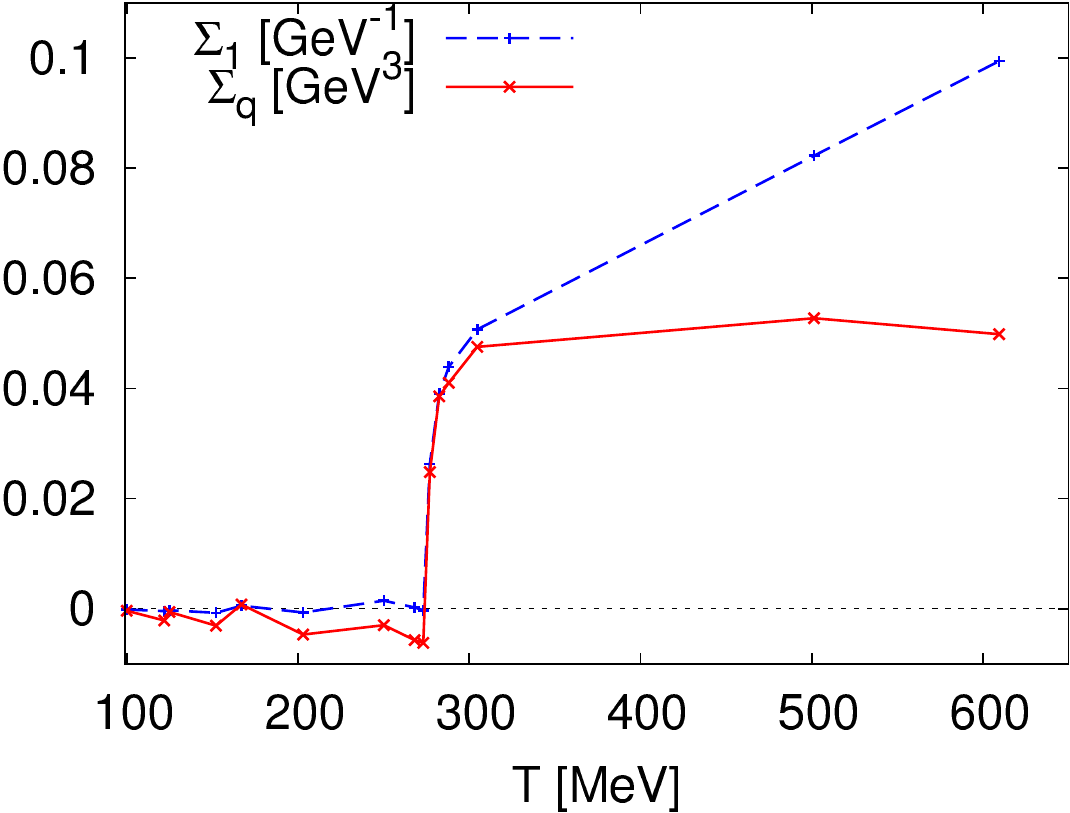}}
  \caption{\label{fig:dse_dual_condensate} Left panel: Quark
    $\langle\bar\psi\psi\rangle_\varphi$ and dual condensate
    $\Sigma_\varphi^{(q)}$, Eq.~\eqref{eq:order_parameter_quark}, as a
    function of the $U(1)$-valued boundary conditions in the chiral
    limit for temperatures above and below the critical temperature
    $T_c=277$ MeV (dashed $T=273$ MeV, solid $T=283$ MeV).  Right
    panel: Order parameters $\Sigma_1$, Eq.~\eqref{eq:dualcond}, and
    $\Sigma_q$, Eq.~(\ref{eq:order_parameter_quark}) as a function of
    the temperature.}
\end{figure*}

To confirm that Eq.~\eqref{eq:order_parameter_quark} defines an order
parameter for the center phase transition in QCD, we present results
for $\Sigma_\varphi^{(q)}$ and $\Sigma_q$ from a self-consistent
treatment of the quark propagator DSE in an analogous setup to
Ref.~\cite{Fischer:2010fx},
cf.\ Refs.~\cite{Fischer:2009wc,Fischer:2009gk}.  In the left panel of
Fig.~\ref{fig:dse_dual_condensate}, the quark condensates
$\langle\bar\psi\psi\rangle_\varphi$ and $\Sigma_\varphi^{(q)}$,
defined in Eq.~(\ref{eq:order_parameter_quark}), are plotted as a
function of the boundary angle $\varphi$. Both quantities show a
similar qualitative behaviour, in particular they are periodic in
$\varphi \in [0,2\pi[$ and symmetric under
$\varphi\rightarrow -\varphi$. Below the critical temperature $T_c$
(dashed lines in the figure), the overall values are non-vanishing and
positive, whereas a sudden drop occurs above $T_c$ (solid
lines). Additionally a flat plateau is formed near antiperiodic
boundary conditions in the chiral limit.  The corresponding order
parameters $\Sigma_1$ and $\Sigma_q$ are shown as a function of
temperature in the right panel of Fig.~\ref{fig:dse_dual_condensate}.
Similar to scalar QCD, artefacts due to the statistical error in
the lattice simulations as well as a dependence on the vertex
model are visible below $T_c$. In fact, we found a stronger dependence
on the vertex model parameter than in the scalar case. This can be explained 
by the small mass of the quark field in comparison to the scalar theory,
since the quark-gluon interaction shows a more pronounced nonperturbative
momentum dependence in the case of small quark masses \cite{Williams:2014iea}.
It is also already clear from purely theoretical considerations that the transition
will only become more pronounced and stable against non-perturbative effects in the
quark-gluon interaction with growing quark mass. With growing mass the quarks turn into
static fundamental sources, whose interaction with the gauge field is
determined by the perturbative quark-gluon vertex strength.
The behaviour of
$\Sigma_q$ below the phase transition could be improved by varying the
model parameters of the quark-gluon vertex. This relatively strong
parameter dependence motivates a more detailed investigation of the
vertex at finite temperature, cf.\ \cite{Huber:2011xc,Hopfer:2013via}.  Above the
critical temperature, $\Sigma_q$ seems to flatten out towards higher
temperatures, which is in contrast to the dual quark condensate
$\Sigma_1$ that shows an almost linear growth in the broken center
symmetry phase.

\section{Summary and conclusions}
\label{sec:summary}

In this letter we have presented novel order parameters for the
confinement-deconfinement phase transition of quenched QCD and scalar
QCD.  These order parameters are sensitive to the spontaneous breaking
of center symmetry, which can be related to confinement in terms of
the free energy of quarks and scalars. Similar to the well-known dual
condensate, we have defined them as the first Fourier series
coefficient of generalized condensates that are determined from matter
propagators with $U(1)$-valued boundary conditions.  In particular, we
introduced a new order parameter for QCD that is based on a quark
propagator dependent condensate, which remains finite even in the case
of non-vanishing current quark masses.

These order parameters are easily accessible with functional methods,
which are usually based on the expansion of the generating functionals
in terms of correlators. Both order parameters rely basically on the
$L^2$-norm which guarantees excellent convergence properties. We
confirmed their validity and accessibility in explicit numerical
calculations of the matter propagators by solving the corresponding
(truncated) Dyson-Schwinger equations. In particular, we find that
even in the case of very heavy scalar matter, where the propagator
itself shows no visible signature of a phase transition, the
discontinuity related to the phase transition can be extracted.
Therefore, these order parameters establish fundamentally charged
static scalars, in particular in terms of their propagators, as a
probe for color confinement.

\subsection*{Acknowledgments}
We thank F.~Bruckmann, C.S.~Fischer, L.~Fister, M.~Huber, A.~Maas,
V.~Mader, J.~M.~Pawlowski, L.~von Smekal and A.~Windisch for valuable
discussions.
M.M. acknowledges support from the FWF through DK-W1203-N16 and
Erwin-Schr\"odinger-Stipendium No. J3507-N27, the Helmholtz Alliance
HA216/EMMI, the BMBF grant OSPL2VHCTG, the grant 
ERC-AdG-290623, the DFG grant MI 2240/1-1 and the U.S. 
Department of Energy under contract de-sc0012704.
M.H. acknowledges support from the FWF through DK-W1203-N16.
B.-J.S. acknowledges support by the FWF grant P24780-N27 and by the
German Federal Ministry of Education and Research (BMBF) under
Contract No. 05P15RGFCA and the Helmholtz International Center for
FAIR within the LOEWE program of the State of Hesse.


\begin{thebibliography}{79}
\expandafter\ifx\csname natexlab\endcsname\relax\def\natexlab#1{#1}\fi
\expandafter\ifx\csname bibnamefont\endcsname\relax
  \def\bibnamefont#1{#1}\fi
\expandafter\ifx\csname bibfnamefont\endcsname\relax
  \def\bibfnamefont#1{#1}\fi
\expandafter\ifx\csname citenamefont\endcsname\relax
  \def\citenamefont#1{#1}\fi
\providecommand{\bibinfo}[2]{#2}
\providecommand{\eprint}[2][]{\url{#2}}

\bibitem{PhysRevD.24.450}
\bibinfo{author}{\bibfnamefont{L.~D.} \bibnamefont{McLerran}} \bibnamefont{and}
  \bibinfo{author}{\bibfnamefont{B.}~\bibnamefont{Svetitsky}},
  \bibinfo{journal}{Phys. Rev. D} \textbf{\bibinfo{volume}{24}},
  \bibinfo{pages}{450} (\bibinfo{year}{1981}).

\bibitem{Pepe:2006er}
\bibinfo{author}{\bibfnamefont{M.}~\bibnamefont{Pepe}} \bibnamefont{and}
  \bibinfo{author}{\bibfnamefont{U.-J.} \bibnamefont{Wiese}},
  \bibinfo{journal}{Nucl.Phys.} \textbf{\bibinfo{volume}{B768}},
  \bibinfo{pages}{21} (\bibinfo{year}{2007}).

\bibitem{Greensite:2006sm}
\bibinfo{author}{\bibfnamefont{J.}~\bibnamefont{Greensite}},
  \bibinfo{author}{\bibfnamefont{K.}~\bibnamefont{Langfeld}},
  \bibinfo{author}{\bibfnamefont{S.}~\bibnamefont{Olejnik}},
  \bibinfo{author}{\bibfnamefont{H.}~\bibnamefont{Reinhardt}},
  \bibnamefont{and} \bibinfo{author}{\bibfnamefont{T.}~\bibnamefont{Tok}},
  \bibinfo{journal}{Phys.Rev.} \textbf{\bibinfo{volume}{D75}},
  \bibinfo{pages}{034501} (\bibinfo{year}{2007}).

\bibitem{Cossu:2007dk}
\bibinfo{author}{\bibfnamefont{G.}~\bibnamefont{Cossu}},
  \bibinfo{author}{\bibfnamefont{M.}~\bibnamefont{D'Elia}},
  \bibinfo{author}{\bibfnamefont{A.}~\bibnamefont{Di~Giacomo}},
  \bibinfo{author}{\bibfnamefont{B.}~\bibnamefont{Lucini}}, \bibnamefont{and}
  \bibinfo{author}{\bibfnamefont{C.}~\bibnamefont{Pica}},
  \bibinfo{journal}{JHEP} \textbf{\bibinfo{volume}{0710}}, \bibinfo{pages}{100}
  (\bibinfo{year}{2007}).

\bibitem{Danzer:2008bk}
\bibinfo{author}{\bibfnamefont{J.}~\bibnamefont{Danzer}},
  \bibinfo{author}{\bibfnamefont{C.}~\bibnamefont{Gattringer}},
  \bibnamefont{and} \bibinfo{author}{\bibfnamefont{A.}~\bibnamefont{Maas}},
  \bibinfo{journal}{JHEP} \textbf{\bibinfo{volume}{0901}}, \bibinfo{pages}{024}
  (\bibinfo{year}{2009}).

\bibitem{Maas:2012wr}
\bibinfo{author}{\bibfnamefont{A.}~\bibnamefont{Maas}},
  \bibinfo{author}{\bibfnamefont{L.}~\bibnamefont{von Smekal}},
  \bibinfo{author}{\bibfnamefont{B.~H.} \bibnamefont{Wellegehausen}},
  \bibnamefont{and} \bibinfo{author}{\bibfnamefont{A.}~\bibnamefont{Wipf}},
  \bibinfo{journal}{Phys.Rev.} \textbf{\bibinfo{volume}{D86}},
  \bibinfo{pages}{111901} (\bibinfo{year}{2012}).

\bibitem{Maas:2012ts}
\bibinfo{author}{\bibfnamefont{A.}~\bibnamefont{Maas}} \bibnamefont{and}
  \bibinfo{author}{\bibfnamefont{B.~H.} \bibnamefont{Wellegehausen}},
  \bibinfo{journal}{PoS} \textbf{\bibinfo{volume}{LATTICE2012}},
  \bibinfo{pages}{080} (\bibinfo{year}{2012}).

\bibitem{Braun:2007bx}
\bibinfo{author}{\bibfnamefont{J.}~\bibnamefont{Braun}},
  \bibinfo{author}{\bibfnamefont{H.}~\bibnamefont{Gies}}, \bibnamefont{and}
  \bibinfo{author}{\bibfnamefont{J.~M.} \bibnamefont{Pawlowski}},
  \bibinfo{journal}{Phys.Lett.} \textbf{\bibinfo{volume}{B684}},
  \bibinfo{pages}{262} (\bibinfo{year}{2010}).

\bibitem{Fister:2013bh}
\bibinfo{author}{\bibfnamefont{L.}~\bibnamefont{Fister}} \bibnamefont{and}
  \bibinfo{author}{\bibfnamefont{J.~M.} \bibnamefont{Pawlowski}},
  \bibinfo{journal}{Phys.Rev.} \textbf{\bibinfo{volume}{D88}},
  \bibinfo{pages}{045010} (\bibinfo{year}{2013}).

\bibitem{Reinhardt:2016xci}
\bibinfo{author}{\bibfnamefont{H.}~\bibnamefont{Reinhardt}},
  \bibinfo{journal}{Phys. Rev.} \textbf{\bibinfo{volume}{D94}},
  \bibinfo{eid}{045016} (\bibinfo{year}{2016}).

\bibitem{Lucini:2012gg}
\bibinfo{author}{\bibfnamefont{B.}~\bibnamefont{Lucini}} \bibnamefont{and}
  \bibinfo{author}{\bibfnamefont{M.}~\bibnamefont{Panero}},
  \bibinfo{journal}{Phys. Rept.} \textbf{\bibinfo{volume}{526}},
  \bibinfo{pages}{93} (\bibinfo{year}{2013}).

\bibitem{Alkofer:2000wg}
\bibinfo{author}{\bibfnamefont{R.}~\bibnamefont{Alkofer}} \bibnamefont{and}
  \bibinfo{author}{\bibfnamefont{L.}~\bibnamefont{von Smekal}},
  \bibinfo{journal}{Phys.Rept.} \textbf{\bibinfo{volume}{353}},
  \bibinfo{pages}{281} (\bibinfo{year}{2001}).

\bibitem{Roberts:2000aa}
\bibinfo{author}{\bibfnamefont{C.~D.} \bibnamefont{Roberts}} \bibnamefont{and}
  \bibinfo{author}{\bibfnamefont{S.~M.} \bibnamefont{Schmidt}},
  \bibinfo{journal}{Prog. Part. Nucl. Phys.} \textbf{\bibinfo{volume}{45}},
  \bibinfo{pages}{S1} (\bibinfo{year}{2000}).

\bibitem{Fischer:2006ub}
\bibinfo{author}{\bibfnamefont{C.~S.} \bibnamefont{Fischer}},
  \bibinfo{journal}{J.Phys.} \textbf{\bibinfo{volume}{G32}},
  \bibinfo{pages}{R253} (\bibinfo{year}{2006}).

\bibitem{Aguilar:2008xm}
\bibinfo{author}{\bibfnamefont{A.~C.} \bibnamefont{Aguilar}},
  \bibinfo{author}{\bibfnamefont{D.}~\bibnamefont{Binosi}}, \bibnamefont{and}
  \bibinfo{author}{\bibfnamefont{J.}~\bibnamefont{Papavassiliou}},
  \bibinfo{journal}{Phys. Rev.} \textbf{\bibinfo{volume}{D78}},
  \bibinfo{pages}{025010} (\bibinfo{year}{2008}).

\bibitem{Fischer:2008uz}
\bibinfo{author}{\bibfnamefont{C.~S.} \bibnamefont{Fischer}},
  \bibinfo{author}{\bibfnamefont{A.}~\bibnamefont{Maas}}, \bibnamefont{and}
  \bibinfo{author}{\bibfnamefont{J.~M.} \bibnamefont{Pawlowski}},
  \bibinfo{journal}{Annals Phys.} \textbf{\bibinfo{volume}{324}},
  \bibinfo{pages}{2408} (\bibinfo{year}{2009}).

\bibitem{Binosi:2009qm}
\bibinfo{author}{\bibfnamefont{D.}~\bibnamefont{Binosi}} \bibnamefont{and}
  \bibinfo{author}{\bibfnamefont{J.}~\bibnamefont{Papavassiliou}},
  \bibinfo{journal}{Phys. Rept.} \textbf{\bibinfo{volume}{479}},
  \bibinfo{pages}{1} (\bibinfo{year}{2009}).

\bibitem{Maas:2011se}
\bibinfo{author}{\bibfnamefont{A.}~\bibnamefont{Maas}},
  \bibinfo{journal}{Phys.Rept.} \textbf{\bibinfo{volume}{524}},
  \bibinfo{pages}{203} (\bibinfo{year}{2013}{\natexlab{a}}).

\bibitem{Boucaud:2011ug}
\bibinfo{author}{\bibfnamefont{P.}~\bibnamefont{Boucaud}},
  \bibinfo{author}{\bibfnamefont{J.~P.} \bibnamefont{Leroy}},
  \bibinfo{author}{\bibfnamefont{A.~L.} \bibnamefont{Yaouanc}},
  \bibinfo{author}{\bibfnamefont{J.}~\bibnamefont{Micheli}},
  \bibinfo{author}{\bibfnamefont{O.}~\bibnamefont{Pene}}, \bibnamefont{and}
  \bibinfo{author}{\bibfnamefont{J.}~\bibnamefont{Rodriguez-Quintero}},
  \bibinfo{journal}{Few Body Syst.} \textbf{\bibinfo{volume}{53}},
  \bibinfo{pages}{387} (\bibinfo{year}{2012}).

\bibitem{Eichmann:2016yit}
\bibinfo{author}{\bibfnamefont{G.}~\bibnamefont{Eichmann}},
  \bibinfo{author}{\bibfnamefont{H.}~\bibnamefont{Sanchis-Alepuz}},
  \bibinfo{author}{\bibfnamefont{R.}~\bibnamefont{Williams}},
  \bibinfo{author}{\bibfnamefont{R.}~\bibnamefont{Alkofer}}, \bibnamefont{and}
  \bibinfo{author}{\bibfnamefont{C.~S.} \bibnamefont{Fischer}},
  \bibinfo{journal}{Prog. Part. Nucl. Phys.} \textbf{\bibinfo{volume}{91}},
  \bibinfo{pages}{1} (\bibinfo{year}{2016}).

\bibitem{Litim:1998nf}
\bibinfo{author}{\bibfnamefont{D.~F.} \bibnamefont{Litim}} \bibnamefont{and}
  \bibinfo{author}{\bibfnamefont{J.~M.} \bibnamefont{Pawlowski}}, in
  \emph{\bibinfo{booktitle}{{The exact renormalization group}}}
  (\bibinfo{address}{Faro, Portugal}, \bibinfo{year}{1998}), pp.
  \bibinfo{pages}{168--185}, \href{http://arxiv.org/abs/hep-th/9901063}{{\tt
  arXiv:hep-th/9901063 [hep-th]}}, 

\bibitem{Berges:2000ew}
\bibinfo{author}{\bibfnamefont{J.}~\bibnamefont{Berges}},
  \bibinfo{author}{\bibfnamefont{N.}~\bibnamefont{Tetradis}}, \bibnamefont{and}
  \bibinfo{author}{\bibfnamefont{C.}~\bibnamefont{Wetterich}},
  \bibinfo{journal}{Phys. Rept.} \textbf{\bibinfo{volume}{363}},
  \bibinfo{pages}{223} (\bibinfo{year}{2002}).

\bibitem{Pawlowski:2005xe}
\bibinfo{author}{\bibfnamefont{J.~M.} \bibnamefont{Pawlowski}},
  \bibinfo{journal}{Annals Phys.} \textbf{\bibinfo{volume}{322}},
  \bibinfo{pages}{2831} (\bibinfo{year}{2007}).

\bibitem{Gies:2006wv}
\bibinfo{author}{\bibfnamefont{H.}~\bibnamefont{Gies}}, \bibinfo{journal}{Lect.
  Notes Phys.} \textbf{\bibinfo{volume}{852}}, \bibinfo{pages}{287}
  (\bibinfo{year}{2012}).

\bibitem{Schaefer:2006sr}
\bibinfo{author}{\bibfnamefont{B.-J.} \bibnamefont{Schaefer}} \bibnamefont{and}
  \bibinfo{author}{\bibfnamefont{J.}~\bibnamefont{Wambach}},
  \bibinfo{journal}{Phys. Part. Nucl.} \textbf{\bibinfo{volume}{39}},
  \bibinfo{pages}{1025} (\bibinfo{year}{2008}).

\bibitem{Rosten:2010vm}
\bibinfo{author}{\bibfnamefont{O.~J.} \bibnamefont{Rosten}},
  \bibinfo{journal}{Phys. Rept.} \textbf{\bibinfo{volume}{511}},
  \bibinfo{pages}{177} (\bibinfo{year}{2012}).

\bibitem{Braun:2011pp}
\bibinfo{author}{\bibfnamefont{J.}~\bibnamefont{Braun}}, \bibinfo{journal}{J.
  Phys.} \textbf{\bibinfo{volume}{G39}}, \bibinfo{pages}{033001}
  (\bibinfo{year}{2012}).

\bibitem{vonSmekal:2012vx}
\bibinfo{author}{\bibfnamefont{L.}~\bibnamefont{von Smekal}},
  \bibinfo{journal}{Nucl. Phys. Proc. Suppl.} \textbf{\bibinfo{volume}{228}},
  \bibinfo{pages}{179} (\bibinfo{year}{2012}).

\bibitem{Marhauser:2008fz}
\bibinfo{author}{\bibfnamefont{F.}~\bibnamefont{Marhauser}} \bibnamefont{and}
  \bibinfo{author}{\bibfnamefont{J.~M.} \bibnamefont{Pawlowski}},
  \href{http://arxiv.org/abs/0812.1144}{{\tt arXiv:0812.1144 [hep-ph]}}.

\bibitem{Fischer:2009wc}
\bibinfo{author}{\bibfnamefont{C.~S.} \bibnamefont{Fischer}},
  \bibinfo{journal}{Phys.Rev.Lett.} \textbf{\bibinfo{volume}{103}},
  \bibinfo{pages}{052003} (\bibinfo{year}{2009}).

\bibitem{Braun:2009gm}
\bibinfo{author}{\bibfnamefont{J.}~\bibnamefont{Braun}},
  \bibinfo{author}{\bibfnamefont{L.~M.} \bibnamefont{Haas}},
  \bibinfo{author}{\bibfnamefont{F.}~\bibnamefont{Marhauser}},
  \bibnamefont{and} \bibinfo{author}{\bibfnamefont{J.~M.}
  \bibnamefont{Pawlowski}}, \bibinfo{journal}{Phys.Rev.Lett.}
  \textbf{\bibinfo{volume}{106}}, \bibinfo{pages}{022002}
  (\bibinfo{year}{2011}).

\bibitem{Fischer:2009gk}
\bibinfo{author}{\bibfnamefont{C.~S.} \bibnamefont{Fischer}} \bibnamefont{and}
  \bibinfo{author}{\bibfnamefont{J.~A.} \bibnamefont{Mueller}},
  \bibinfo{journal}{Phys.Rev.} \textbf{\bibinfo{volume}{D80}},
  \bibinfo{pages}{074029} (\bibinfo{year}{2009}).

\bibitem{Fischer:2010fx}
\bibinfo{author}{\bibfnamefont{C.~S.} \bibnamefont{Fischer}},
  \bibinfo{author}{\bibfnamefont{A.}~\bibnamefont{Maas}}, \bibnamefont{and}
  \bibinfo{author}{\bibfnamefont{J.~A.} \bibnamefont{Mueller}},
  \bibinfo{journal}{Eur.Phys.J.} \textbf{\bibinfo{volume}{C68}},
  \bibinfo{pages}{165} (\bibinfo{year}{2010}).

\bibitem{Fischer:2011mz}
\bibinfo{author}{\bibfnamefont{C.~S.} \bibnamefont{Fischer}},
  \bibinfo{author}{\bibfnamefont{J.}~\bibnamefont{Luecker}}, \bibnamefont{and}
  \bibinfo{author}{\bibfnamefont{J.~A.} \bibnamefont{Mueller}},
  \bibinfo{journal}{Phys.Lett.} \textbf{\bibinfo{volume}{B702}},
  \bibinfo{pages}{438} (\bibinfo{year}{2011}).

\bibitem{Fischer:2012vc}
\bibinfo{author}{\bibfnamefont{C.~S.} \bibnamefont{Fischer}} \bibnamefont{and}
  \bibinfo{author}{\bibfnamefont{J.}~\bibnamefont{Luecker}},
  \bibinfo{journal}{Phys.Lett.} \textbf{\bibinfo{volume}{B718}},
  \bibinfo{pages}{1036} (\bibinfo{year}{2013}).

\bibitem{Haas:2013qwp}
\bibinfo{author}{\bibfnamefont{L.~M.} \bibnamefont{Haas}},
  \bibinfo{author}{\bibfnamefont{R.}~\bibnamefont{Stiele}},
  \bibinfo{author}{\bibfnamefont{J.}~\bibnamefont{Braun}},
  \bibinfo{author}{\bibfnamefont{J.~M.} \bibnamefont{Pawlowski}},
  \bibnamefont{and}
  \bibinfo{author}{\bibfnamefont{J.}~\bibnamefont{Schaffner-Bielich}},
  \bibinfo{journal}{Phys.Rev.} \textbf{\bibinfo{volume}{D87}},
  \bibinfo{pages}{076004} (\bibinfo{year}{2013}).

\bibitem{Fischer:2013eca}
\bibinfo{author}{\bibfnamefont{C.~S.} \bibnamefont{Fischer}},
  \bibinfo{author}{\bibfnamefont{L.}~\bibnamefont{Fister}},
  \bibinfo{author}{\bibfnamefont{J.}~\bibnamefont{Luecker}}, \bibnamefont{and}
  \bibinfo{author}{\bibfnamefont{J.~M.} \bibnamefont{Pawlowski}},
  \bibinfo{journal}{Physics Letters B} \textbf{\bibinfo{volume}{732}},
  \bibinfo{pages}{273} (\bibinfo{year}{2014}{\natexlab{a}}).

\bibitem{Fischer:2014ata}
\bibinfo{author}{\bibfnamefont{C.~S.} \bibnamefont{Fischer}},
  \bibinfo{author}{\bibfnamefont{J.}~\bibnamefont{Luecker}}, \bibnamefont{and}
  \bibinfo{author}{\bibfnamefont{C.~A.} \bibnamefont{Welzbacher}},
  \bibinfo{journal}{Phys. Rev.} \textbf{\bibinfo{volume}{D90}},
  \bibinfo{pages}{034022} (\bibinfo{year}{2014}{\natexlab{b}}).

\bibitem{Quandt:2015aaa}
\bibinfo{author}{\bibfnamefont{M.}~\bibnamefont{Quandt}} \bibnamefont{and}
  \bibinfo{author}{\bibfnamefont{H.}~\bibnamefont{Reinhardt}},
  \bibinfo{journal}{Phys. Rev.} \textbf{\bibinfo{volume}{D92}},
  \bibinfo{pages}{025051} (\bibinfo{year}{2015}).

\bibitem{Reinosa:2016iml}
\bibinfo{author}{\bibfnamefont{U.}~\bibnamefont{Reinosa}},
  \bibinfo{author}{\bibfnamefont{J.}~\bibnamefont{Serreau}},
  \bibinfo{author}{\bibfnamefont{M.}~\bibnamefont{Tissier}}, \bibnamefont{and}
  \bibinfo{author}{\bibfnamefont{A.}~\bibnamefont{Tresmontant}},
  \bibinfo{journal}{Phys. Rev.} \textbf{\bibinfo{volume}{D95}},
  \bibinfo{pages}{045014} (\bibinfo{year}{2017}).

\bibitem{Huber:2016xbs}
\bibinfo{author}{\bibfnamefont{M.~Q.} \bibnamefont{Huber}},
  \bibinfo{journal}{EPJ Web Conf.} \textbf{\bibinfo{volume}{137}},
  \bibinfo{pages}{07009} (\bibinfo{year}{2017}).

\bibitem{Cyrol:2017qkl}
\bibinfo{author}{\bibfnamefont{A.~K.} \bibnamefont{Cyrol}},
  \bibinfo{author}{\bibfnamefont{M.}~\bibnamefont{Mitter}},
  \bibinfo{author}{\bibfnamefont{J.~M.} \bibnamefont{Pawlowski}},
  \bibnamefont{and}
  \bibinfo{author}{\bibfnamefont{N.}~\bibnamefont{Strodthoff}},
  \href{http://arxiv.org/abs/1708.03482}{{\tt arXiv:1708.03482 [hep-ph]}}.

\bibitem{Herbst:2015ona}
\bibinfo{author}{\bibfnamefont{T.~K.} \bibnamefont{Herbst}},
  \bibinfo{author}{\bibfnamefont{J.}~\bibnamefont{Luecker}}, \bibnamefont{and}
  \bibinfo{author}{\bibfnamefont{J.~M.} \bibnamefont{Pawlowski}},
  \href{http://arxiv.org/abs/1510.03830}{{\tt arXiv:1510.03830 [hep-ph]}}.

\bibitem{Gattringer:2006ci}
\bibinfo{author}{\bibfnamefont{C.}~\bibnamefont{Gattringer}},
  \bibinfo{journal}{Phys.Rev.Lett.} \textbf{\bibinfo{volume}{97}},
  \bibinfo{pages}{032003} (\bibinfo{year}{2006}).

\bibitem{Bruckmann:2006kx}
\bibinfo{author}{\bibfnamefont{F.}~\bibnamefont{Bruckmann}},
  \bibinfo{author}{\bibfnamefont{C.}~\bibnamefont{Gattringer}},
  \bibnamefont{and} \bibinfo{author}{\bibfnamefont{C.}~\bibnamefont{Hagen}},
  \bibinfo{journal}{Phys.Lett.} \textbf{\bibinfo{volume}{B647}},
  \bibinfo{pages}{56} (\bibinfo{year}{2007}).

\bibitem{Bilgici:2008qy}
\bibinfo{author}{\bibfnamefont{E.}~\bibnamefont{Bilgici}},
  \bibinfo{author}{\bibfnamefont{F.}~\bibnamefont{Bruckmann}},
  \bibinfo{author}{\bibfnamefont{C.}~\bibnamefont{Gattringer}},
  \bibnamefont{and} \bibinfo{author}{\bibfnamefont{C.}~\bibnamefont{Hagen}},
  \bibinfo{journal}{Phys.Rev.} \textbf{\bibinfo{volume}{D77}},
  \bibinfo{pages}{094007} (\bibinfo{year}{2008}).

\bibitem{Synatschke:2007bz}
\bibinfo{author}{\bibfnamefont{F.}~\bibnamefont{Synatschke}},
  \bibinfo{author}{\bibfnamefont{A.}~\bibnamefont{Wipf}}, \bibnamefont{and}
  \bibinfo{author}{\bibfnamefont{C.}~\bibnamefont{Wozar}},
  \bibinfo{journal}{Phys.Rev.} \textbf{\bibinfo{volume}{D75}},
  \bibinfo{pages}{114003} (\bibinfo{year}{2007}).

\bibitem{Synatschke:2008yt}
\bibinfo{author}{\bibfnamefont{F.}~\bibnamefont{Synatschke}},
  \bibinfo{author}{\bibfnamefont{A.}~\bibnamefont{Wipf}}, \bibnamefont{and}
  \bibinfo{author}{\bibfnamefont{K.}~\bibnamefont{Langfeld}},
  \bibinfo{journal}{Phys.Rev.} \textbf{\bibinfo{volume}{D77}},
  \bibinfo{pages}{114018} (\bibinfo{year}{2008}).

\bibitem{Fister:2010ah}
\bibinfo{author}{\bibfnamefont{L.}~\bibnamefont{Fister}}, Ph.D. thesis,
  \bibinfo{school}{Heidelberg U.} (\bibinfo{year}{2009}),
  \href{http://arxiv.org/abs/1002.1649}{{\tt arXiv:1002.1649 [hep-th]}},

\bibitem{Fister:2010yw}
\bibinfo{author}{\bibfnamefont{L.}~\bibnamefont{Fister}},
  \bibinfo{author}{\bibfnamefont{R.}~\bibnamefont{Alkofer}}, \bibnamefont{and}
  \bibinfo{author}{\bibfnamefont{K.}~\bibnamefont{Schwenzer}},
  \bibinfo{journal}{Phys.Lett.} \textbf{\bibinfo{volume}{B688}},
  \bibinfo{pages}{237} (\bibinfo{year}{2010}).

\bibitem{Macher:2011ys}
\bibinfo{author}{\bibfnamefont{V.}~\bibnamefont{Macher}},
  \bibinfo{author}{\bibfnamefont{A.}~\bibnamefont{Maas}}, \bibnamefont{and}
  \bibinfo{author}{\bibfnamefont{R.}~\bibnamefont{Alkofer}},
  \bibinfo{journal}{Int.J.Mod.Phys.} \textbf{\bibinfo{volume}{A27}},
  \bibinfo{pages}{1250098} (\bibinfo{year}{2012}).

\bibitem{Huber:2011xc}
\bibinfo{author}{\bibfnamefont{M.~Q.} \bibnamefont{Huber}} \bibnamefont{and}
  \bibinfo{author}{\bibfnamefont{M.}~\bibnamefont{Mitter}},
  \bibinfo{journal}{Comput.Phys.Commun.} \textbf{\bibinfo{volume}{183}},
  \bibinfo{pages}{2441} (\bibinfo{year}{2012}).

\bibitem{Maas:2012tj}
\bibinfo{author}{\bibfnamefont{A.}~\bibnamefont{Maas}}, \bibinfo{journal}{Mod.
  Phys. Lett.} \textbf{\bibinfo{volume}{A28}}, \bibinfo{pages}{1350103}
  (\bibinfo{year}{2013}{\natexlab{b}}).

\bibitem{Mitter:2012tpo}
\bibinfo{author}{\bibfnamefont{M.}~\bibnamefont{Mitter}}, Ph.D. thesis,
  \bibinfo{school}{Graz U.} (\bibinfo{year}{2012}), 
  INSPIRE-1417170;

\bibitem{Hopfer:2012qr}
\bibinfo{author}{\bibfnamefont{M.}~\bibnamefont{Hopfer}},
  \bibinfo{author}{\bibfnamefont{M.}~\bibnamefont{Mitter}},
  \bibinfo{author}{\bibfnamefont{B.-J.} \bibnamefont{Schaefer}},
  \bibnamefont{and} \bibinfo{author}{\bibfnamefont{R.}~\bibnamefont{Alkofer}},
  \bibinfo{journal}{Acta Phys.Polon.Supp.} \textbf{\bibinfo{volume}{6}},
  \bibinfo{pages}{353} (\bibinfo{year}{2013}).

\bibitem{Mitter:2013me}
\bibinfo{author}{\bibfnamefont{M.}~\bibnamefont{Mitter}},
  \bibinfo{author}{\bibfnamefont{M.}~\bibnamefont{Hopfer}},
  \bibinfo{author}{\bibfnamefont{B.-J.} \bibnamefont{Schaefer}},
  \bibnamefont{and} \bibinfo{author}{\bibfnamefont{R.}~\bibnamefont{Alkofer}},
  \bibinfo{journal}{PoS} \textbf{\bibinfo{volume}{ConfinementX}},
  \bibinfo{pages}{195} (\bibinfo{year}{2012}).

\bibitem{Hopfer:2013via}
\bibinfo{author}{\bibfnamefont{M.}~\bibnamefont{Hopfer}} \bibnamefont{and}
  \bibinfo{author}{\bibfnamefont{R.}~\bibnamefont{Alkofer}},
  \bibinfo{journal}{Acta Phys.Polon.Supp.} \textbf{\bibinfo{volume}{6}},
  \bibinfo{pages}{929} (\bibinfo{year}{2013}).

\bibitem{Maas:2013aia}
\bibinfo{author}{\bibfnamefont{A.}~\bibnamefont{Maas}} \bibnamefont{and}
  \bibinfo{author}{\bibfnamefont{T.}~\bibnamefont{Mufti}},
  \bibinfo{journal}{JHEP} \textbf{\bibinfo{volume}{04}}, \bibinfo{pages}{006}
  (\bibinfo{year}{2014}).

\bibitem{Maas:2014pba}
\bibinfo{author}{\bibfnamefont{A.}~\bibnamefont{Maas}} \bibnamefont{and}
  \bibinfo{author}{\bibfnamefont{T.}~\bibnamefont{Mufti}},
  \bibinfo{journal}{Phys. Rev.} \textbf{\bibinfo{volume}{D91}},
  \bibinfo{pages}{113011} (\bibinfo{year}{2015}).

\bibitem{Hopfer:2014szm}
\bibinfo{author}{\bibfnamefont{M.}~\bibnamefont{Hopfer}}, Ph.D. thesis,
  \bibinfo{school}{Graz U.} (\bibinfo{year}{2014}), 
  INSPIRE-1417166;

\bibitem{Maas:2016edk}
\bibinfo{author}{\bibfnamefont{A.}~\bibnamefont{Maas}}, \bibinfo{journal}{Eur.
  Phys. J.} \textbf{\bibinfo{volume}{C76}}, \bibinfo{pages}{366}
  (\bibinfo{year}{2016}).

\bibitem{Maas:2016ngo}
\bibinfo{author}{\bibfnamefont{A.}~\bibnamefont{Maas}} \bibnamefont{and}
  \bibinfo{author}{\bibfnamefont{P.}~\bibnamefont{Torek}},
  \bibinfo{journal}{Phys. Rev.} \textbf{\bibinfo{volume}{D95}},
  \bibinfo{pages}{014501} (\bibinfo{year}{2017}).

\bibitem{Osterwalder:1977pc}
\bibinfo{author}{\bibfnamefont{K.}~\bibnamefont{Osterwalder}} \bibnamefont{and}
  \bibinfo{author}{\bibfnamefont{E.}~\bibnamefont{Seiler}},
  \bibinfo{journal}{Annals Phys.} \textbf{\bibinfo{volume}{110}},
  \bibinfo{pages}{440} (\bibinfo{year}{1978}).

\bibitem{Bertle:2003pj}
\bibinfo{author}{\bibfnamefont{R.}~\bibnamefont{Bertle}},
  \bibinfo{author}{\bibfnamefont{M.}~\bibnamefont{Faber}},
  \bibinfo{author}{\bibfnamefont{J.}~\bibnamefont{Greensite}},
  \bibnamefont{and} \bibinfo{author}{\bibfnamefont{S.}~\bibnamefont{Olejnik}},
  \bibinfo{journal}{Phys.Rev.} \textbf{\bibinfo{volume}{D69}},
  \bibinfo{pages}{014007} (\bibinfo{year}{2004}).

\bibitem{Langfeld:2004vu}
\bibinfo{author}{\bibfnamefont{K.}~\bibnamefont{Langfeld}},
  \bibinfo{journal}{AIP Conf.Proc.} \textbf{\bibinfo{volume}{756}},
  \bibinfo{pages}{133} (\bibinfo{year}{2005}).

\bibitem{Schaden:2013ffa}
\bibinfo{author}{\bibfnamefont{V.}~\bibnamefont{Mader}},
  \bibinfo{author}{\bibfnamefont{M.}~\bibnamefont{Schaden}},
  \bibinfo{author}{\bibfnamefont{D.}~\bibnamefont{Zwanziger}},
  \bibnamefont{and} \bibinfo{author}{\bibfnamefont{R.}~\bibnamefont{Alkofer}},
  \bibinfo{journal}{Eur. Phys. J.} \textbf{\bibinfo{volume}{C74}},
  \bibinfo{pages}{2881} (\bibinfo{year}{2014}).

\bibitem{Alkofer:2008tt}
\bibinfo{author}{\bibfnamefont{R.}~\bibnamefont{Alkofer}},
  \bibinfo{author}{\bibfnamefont{C.~S.} \bibnamefont{Fischer}},
  \bibinfo{author}{\bibfnamefont{F.~J.} \bibnamefont{Llanes-Estrada}},
  \bibnamefont{and}
  \bibinfo{author}{\bibfnamefont{K.}~\bibnamefont{Schwenzer}},
  \bibinfo{journal}{Annals Phys.} \textbf{\bibinfo{volume}{324}},
  \bibinfo{pages}{106} (\bibinfo{year}{2009}).

\bibitem{Williams:2014iea}
\bibinfo{author}{\bibfnamefont{R.}~\bibnamefont{Williams}},
  \bibinfo{journal}{Eur. Phys. J.} \textbf{\bibinfo{volume}{A51}},
  \bibinfo{pages}{57} (\bibinfo{year}{2015}).

\bibitem{Mitter:2014wpa}
\bibinfo{author}{\bibfnamefont{M.}~\bibnamefont{Mitter}},
  \bibinfo{author}{\bibfnamefont{J.~M.} \bibnamefont{Pawlowski}},
  \bibnamefont{and}
  \bibinfo{author}{\bibfnamefont{N.}~\bibnamefont{Strodthoff}},
  \bibinfo{journal}{Phys. Rev.} \textbf{\bibinfo{volume}{D91}},
  \bibinfo{pages}{054035} (\bibinfo{year}{2015}).

\bibitem{Williams:2015cvx}
\bibinfo{author}{\bibfnamefont{R.}~\bibnamefont{Williams}},
  \bibinfo{author}{\bibfnamefont{C.~S.} \bibnamefont{Fischer}},
  \bibnamefont{and} \bibinfo{author}{\bibfnamefont{W.}~\bibnamefont{Heupel}},
  \bibinfo{journal}{Phys. Rev.} \textbf{\bibinfo{volume}{D93}},
  \bibinfo{pages}{034026} (\bibinfo{year}{2016}).

\bibitem{Binosi:2016wcx}
\bibinfo{author}{\bibfnamefont{D.}~\bibnamefont{Binosi}},
  \bibinfo{author}{\bibfnamefont{L.}~\bibnamefont{Chang}},
  \bibinfo{author}{\bibfnamefont{J.}~\bibnamefont{Papavassiliou}},
  \bibinfo{author}{\bibfnamefont{S.-X.} \bibnamefont{Qin}}, \bibnamefont{and}
  \bibinfo{author}{\bibfnamefont{C.~D.} \bibnamefont{Roberts}},
  \bibinfo{journal}{Phys. Rev.} \textbf{\bibinfo{volume}{D95}},
  \bibinfo{pages}{031501} (\bibinfo{year}{2017}).

\bibitem{Aguilar:2016lbe}
\bibinfo{author}{\bibfnamefont{A.~C.} \bibnamefont{Aguilar}},
  \bibinfo{author}{\bibfnamefont{J.~C.} \bibnamefont{Cardona}},
  \bibinfo{author}{\bibfnamefont{M.~N.} \bibnamefont{Ferreira}},
  \bibnamefont{and}
  \bibinfo{author}{\bibfnamefont{J.}~\bibnamefont{Papavassiliou}},
  \bibinfo{journal}{Phys. Rev.} \textbf{\bibinfo{volume}{D96}},
  \bibinfo{pages}{014029} (\bibinfo{year}{2017}).

\bibitem{Cyrol:2017ewj}
\bibinfo{author}{\bibfnamefont{A.~K.} \bibnamefont{Cyrol}},
  \bibinfo{author}{\bibfnamefont{M.}~\bibnamefont{Mitter}},
  \bibinfo{author}{\bibfnamefont{J.~M.} \bibnamefont{Pawlowski}},
  \bibnamefont{and}
  \bibinfo{author}{\bibfnamefont{N.}~\bibnamefont{Strodthoff}},
  \href{http://arxiv.org/abs/1706.06326}{{\tt arXiv:1706.06326 [hep-ph]}}.

\bibitem{Marciano:1977su}
\bibinfo{author}{\bibfnamefont{W.~J.} \bibnamefont{Marciano}} \bibnamefont{and}
  \bibinfo{author}{\bibfnamefont{H.}~\bibnamefont{Pagels}},
  \bibinfo{journal}{Phys.Rept.} \textbf{\bibinfo{volume}{36}},
  \bibinfo{pages}{137} (\bibinfo{year}{1978}).

\bibitem{Taylor:1971ff}
\bibinfo{author}{\bibfnamefont{J.}~\bibnamefont{Taylor}},
  \bibinfo{journal}{Nucl.Phys.} \textbf{\bibinfo{volume}{B33}},
  \bibinfo{pages}{436} (\bibinfo{year}{1971}).

\bibitem{Huber:2014tva}
\bibinfo{author}{\bibfnamefont{M.~Q.} \bibnamefont{Huber}} \bibnamefont{and}
  \bibinfo{author}{\bibfnamefont{L.}~\bibnamefont{von Smekal}},
  \bibinfo{journal}{JHEP} \textbf{\bibinfo{volume}{06}}, \bibinfo{pages}{015}
  (\bibinfo{year}{2014}).

\bibitem{Ball:1980ay}
\bibinfo{author}{\bibfnamefont{J.~S.} \bibnamefont{Ball}} \bibnamefont{and}
  \bibinfo{author}{\bibfnamefont{T.-W.} \bibnamefont{Chiu}},
  \bibinfo{journal}{Phys.Rev.} \textbf{\bibinfo{volume}{D22}},
  \bibinfo{pages}{2542} (\bibinfo{year}{1980}{\natexlab{a}}).

\bibitem{Ball:1980ax}
\bibinfo{author}{\bibfnamefont{J.~S.} \bibnamefont{Ball}} \bibnamefont{and}
  \bibinfo{author}{\bibfnamefont{T.-W.} \bibnamefont{Chiu}},
  \bibinfo{journal}{Phys.Rev.} \textbf{\bibinfo{volume}{D22}},
  \bibinfo{pages}{2550} (\bibinfo{year}{1980}{\natexlab{b}}).

\bibitem{Fischer:2003rp}
\bibinfo{author}{\bibfnamefont{C.~S.} \bibnamefont{Fischer}} \bibnamefont{and}
  \bibinfo{author}{\bibfnamefont{R.}~\bibnamefont{Alkofer}},
  \bibinfo{journal}{Phys.Rev.} \textbf{\bibinfo{volume}{D67}},
  \bibinfo{pages}{094020} (\bibinfo{year}{2003}).

\end{thebibliography}

\end{document}